# A Simplified Analytical Approach for Determining Eclipses of Satellites Occulted by a Celestial Body

Vladislav Zubko[*,1], Andrey Belyaev[1]

[1] *Space Research Institute of the Russian Academy of Sciences, ul. Profsoyuznaya 84/32, Moscow, 117997 Russia*

**Abstract**—The primary objective of this paper is to construct an analytical model for determining the total duration of eclipse events of satellites. The approach assumes that the trace formed in the orbital plane, cutting body's shadow under the classical conical shadow model, can be described as an ellipse. This allows for the derivation of its parameters through the use of classical orbital elements and solar position in the body's centric frame. Consequently, the problem of identifying satellite occultation events is simplified to searching for points of intersection between two ellipses: the satellite orbit and the shadow. The developed model has been demonstrated to be applicable for a wide range of inclinations, with the exception of cases where the orbital plane of the satellite's motion coincides with that of the sun's in the body-centric frame. It is shown that the shadow function constructed under the aforementioned assumptions can be simplified in the cylindrical shadow model. Several simplifications to the proposed model are presented in the study. The paper also considers extending the model to account for non-spherical shapes of celestial bodies, i.e., bodies with oblateness, and presents an algorithm that accounts for changes in eclipse duration due to the heliocentric motion of these celestial bodies. The model has been validated through numerous tests with a numerical algorithm and satellite real data, as well as with other analytical models.

## NOMENCLATURE

$X_\gamma Y$ - unperturbed plane of motion of the planet;

$X_\gamma, Y, Z$ - planet-centric frame originated at planet's center, where $X_\gamma$ is directed to the point of intersection between planet's plane and ecliptic, $Z$ coincides with planet's orbital momentum;

$a_1$ = semimajor axis of satellite orbit, [km];

$a_2, b_2$ = semimajor and semiminor axes of the shadow ellipse, [km];

$e_1$ = eccentricity of satellite orbit;

$e_2$ = eccentricity of the shadow ellipse;

---

[1] *Corresponding author*: Vladislav Zubko is PhD student, researcher at the Space Research Institute (IKI) of the Russian Academy of Sciences (RAS). E-mail: v.zubko@iki.rssi.ru

$E_1$=eccentric anomaly of satellite, [rad];

$H_1$=hyperbolic anomaly of satellite, [rad];

$f_1$ = true anomaly of satellite, [rad];

$i_1$ = inclination of the satellite orbit to the $X_\gamma, Y$ plane, [rad];

$p_1 = a_1(1-e_1^2)$ denotes the semilatus rectum of the satellite orbit, [km];

$r_1 = \frac{p_1}{1+e_1 \cos f_1}$ is the radius of the spacecraft in orbit, [km];

$r_{\pi 1} = a_1(1-e_1)$ is the periapsis radius, [km];

$\mathbf{r}_\odot$ = position of the Sun in the $X_\gamma, Y, Z$ frame, [km];

$\mathbf{r}_\odot^* = -\mathbf{r}_\odot$ = vector that is opposite to the Sun direction in the $X_\gamma, Y, Z$ frame;

$r_\odot = \|\mathbf{r}_\odot\|$ distance between the centre of the Sun and the planet, [km];

$R_\odot, R$ = the average solar and the planetary radii, [km].

$\alpha_p, \alpha_u$ = the penumbra (includes umbra), and umbra half angles, [deg];

$\beta_\odot$ = angle between the $\mathbf{r}_\odot^*$ and tits projection on the orbital plane, i.e. sun-orbital angle, [deg];

$\gamma$ = angle between the satellite position at the shadow boundary and $\mathbf{r}_\odot^*$, [deg];

$\lambda_1 = \Omega_1 + \omega_1$ = longitude of the periapsis in the $X_\gamma, Y, Z$ frame applied when $i_1 = 0\,\deg$, [deg];

$\lambda_\odot$ = right ascension of the Sun in the $X_\gamma, Y, Z$ frame, [deg];

$\mu_{pl}$ = gravitational parameter of the planet, [km$^3$/s$^2$];

$P_{pl}$ = heliocentric orbital period of planet, [s];

$\omega_1$ = argument of periapsis of satellite orbit given in frame $X_\gamma, Y, Z$, [deg];

$\Omega_1$ = longitude of the ascending node of satellite orbit given in the $X_\gamma, Y, Z$ frame, [deg];

$\varpi_\odot$ = angle between projection of $\mathbf{r}_\odot^*$ on the satellite orbital plane and vector directed from planet's centre to the satellite's orbit periapsis, [deg];

$\xi, \eta, \zeta$ are the axes for the orbital frame, where $\xi$ is directed to the periapsis, $\zeta$ lies in the normal direction to the orbital plane and $\eta$ completes the right-handed axes.

**Notice**. symbol "$+/-$" in the subscript denotes positions and angles that correspond to the satellite position at exit/entry from and to the planet shadow, e.g., $\mathbf{r}_+$ is the position of the satellite at the shadow exit point. Conversely, the superscript "$in/sc$" denotes only the intersection points between the shadow of the planet and the satellite orbit.

# INTRODUCTION

The problem of determining intervals of a satellite eclipse duration is important for many applications in many problems of theoretical astronomy and astrodynamics. Accurately measuring the amount of time that a satellite spends in the shadow of a celestial body requires understanding the definition and characteristics of the umbra and penumbra regions of the satellite's orbit [1,2].

Accurate knowledge of crossing the umbra and penumbra regions is essential for determining the current position of satellite, navigation and communication means, and prediction of its position when calculating in the model consisting of sunlight radiation perturbation.

The umbra region is a conical total shadow formed on the opposite side of the celestial body, where solar radiation intensity is close to zero. The penumbra region, on the other hand, is the partial shadow between the umbra and full-light regions, where sunlight is only partially blocked. Generally, the penumbra is considered a region that includes both full and partial shadows, as shown in Fig. 1. Both the penumbra and umbra in the conical model are considered cones with a common rotation axis. The half-angles that determine each of these cones are derived from the geometry in Fig. 1 [3]:

$$\begin{cases} \sin\alpha_u = \dfrac{R_\odot - R}{r_\odot} \\ \sin\alpha_p = \dfrac{R_\odot + R}{r_\odot} \end{cases} \tag{1}$$

The problem derived in this work is an old fundamental problem of astrodynamics that has been researched by many scientists. Analytical solutions have been established in various studies. Studies, such as [4–9], have suggested using quartic equation analytic solutions to determine umbra and penumbra boundaries. However, this method requires checking all four solutions for spurious ones since the quartic equation is the result of squaring the original equation.

In [1], the authors proposed a different approach that avoids solving the original quartic equation for certain partial cases. They compared different approaches and concluded that a numerical solution was the most effective. This alternative and more precise method for determining umbra and penumbra boundaries is crucial for space missions and satellite operations.

Recent research [10] has approached the problem of estimating the time a satellite spends in the shadow of a spherical planet. Assuming the satellite moves in the ecliptic plane and the surface of the shadow forms an ideal circular cone, the paper provides numerical and analytical estimates of the time a satellite spends in the shadow of the Moon and the Earth for cylindrical and conical models of the shadow.

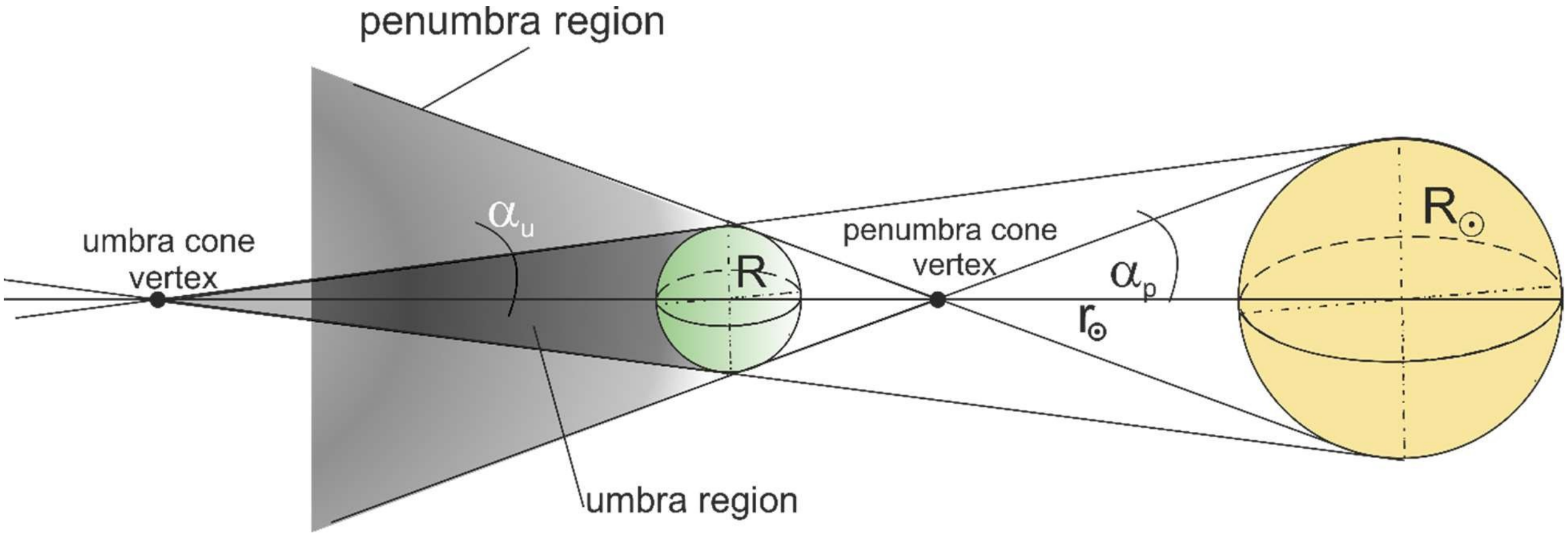

*Figure 1 Geometry of both umbra ($\alpha_u$) and penumbra ($\alpha_p$) shadow cones*

Ref. [11] presents an analytical solution for the conical eclipse of an elliptical orbit with small eccentricity, revealing the effects of eccentricity on the eclipse and challenging the notion of the negligibility of the penumbra arc length. The accuracy of the proposed model is validated through numerical simulation, demonstrating its effectiveness in common situations but highlighting its limitations near the critical orbital sun angle.

There are also analytical methods for determining the entry and exit times of low-Earth orbiting satellites taking into account Earth's oblateness [12]. Their method enable prediction the entry and exit times within approximately 3-4 seconds when compared with truth data from satellites. Additionally, it is worth noting that an approach for the analytical calculation of shadow boundaries was proposed in [13], which is suitable for calculating shadowing in the case of oblateness. In a study [14], it was determined that the Earth's oblateness does not have a notable influence on the accuracy of predicting the entry and exit times of eclipses for satellites in low-Earth orbit. Building upon this research, [15,16] took the oblate-Earth conical model a step further and conducted a comparative analysis between the model, commercial software, and real satellite data.

The study [17] examines the impact of Earth's oblateness on predicting shadow events of a lunar spacecraft caused by Earth's shadow. The modified 'line-of-intersection' method is applied to predict shadow conditions using a spheroidal Earth model, showing significant variations in predicted entry and exit times, as well as durations, for different lunar mission phases. The findings emphasize the importance of considering Earth's oblate shape in predicting shadow events for distant spacecraft and provide insights for missions involving highly elliptical orbits around Earth or travel to the Moon. Another approach for simplified models that can be used in practice for the determination of shadowing time considered oblate Earth conical/cylindrical shadowing models is described in [15,18,19].

Implementation of shadowing techniques for other planet exploration missions was made in [20] to find the eclipse of the Mars orbit. Song and Kim [17] proposed a modification to the line-of-intersection method, incorporating considerations for Earth oblateness, to improve the accuracy of calculating the impact of an eclipse on the Moon's artificial satellite orbit. The authors of [21] used a smoothing technique to reduce the jumping effect caused by entering and exiting the eclipse during the symplectic integration of space debris motion. In [22], the authors discussed the increasing interest in understanding trajectories in cislunar space and the challenges posed by Earth and Moon eclipses. Their study investigates strategies to predict and mitigate eclipses, presenting techniques that leverage trajectory geometries and implement path constraints to avoid penumbra, ultimately demonstrating the utility of synodic resonances for sustaining cislunar operations.

Additionally, the paper [23] is worth highlighting. In this paper author discuss the importance of accurate modelling of solar radiation pressure (SRP) in the precise orbit determination (POD) of GNSS satellites. It compares four different shadow models and assesses their impact on orbit accuracy through validation and range checks. The results demonstrate that considering Earth's oblateness and atmospheric effects, particularly for perturbative accelerations higher than $10^{-10} - 10^{-8}$ m/s$^2$, can significantly improve orbit accuracy by reducing the root mean square (RMS) and improving satellite laser ranging (SLR) validation and intersatellite link (ISL) checks.

The current paper focuses on establishing a universal approach for the construction of a model for satellite eclipse prediction (hereafter, the model will be regarded as the Shadow Elliptic Eclipse Model (SEEM)). The basis for this is the consideration of a generally used model for defining umbra and penumbra regions as cones. In this case, the satellite orbital plane that cuts these cones and the line of intersection that is generated in the orbital plane can be described as a second-order curve that can be either an ellipse, parabola or pair of straight lines. Ergo, for occultation event detection, it is necessary to solve the 4$^{th}$-degree algebraic polynomial to find the points of intersection between the mentioned curves. Consequently, the general solution for either elliptical or hyperbolic orbits of satellites over a wide range of solar-orbital angles can be found.

The main advantage of the presented method is its full analyticity for any $\beta_{\odot}$, and also for any problem of search for occultation events if within this problem geometrical representation is possible, for example, account of compression of a planet, account of refraction of rays by an atmosphere of a planet, and also relativistic effects. An example with the construction of an analytical model when taking into account the compression of the planet will be presented in the paper.

Another advantage of the developed SEEM model is that the derived equations necessary to determine the shadow boundaries in the orbital plane depend entirely on the satellite's orbital (Keplerian) elements and solar longitude. This fact can serve as a reference for the construction of new analytical models, in particular, for vehicles using low thrust spiral trajectory spin-up, as well as for the construction of new analytical models of motion of highly elliptical near-planetary satellites. Additionally, SEEM can be easily integrated into algorithms for searching trajectories of spacecraft flight to other planets with accompanying gravity assist manoeuvres to calculate shadow regions in spacecraft motion within the sphere of influence of a planet.

Additionally, the article studies the extension of the developed model to account for satellites orbiting nonspherical bodies (i.e., considering oblateness) and presents an algorithm for factoring in the effect of planet motion during satellite orbit. The results obtained by solving the problem are significant for accurate and efficient calculations of orbits and are proven to be more efficient in generating quick estimates compared with computationally demanding numerical propagation, thereby providing an advantage in project calculations of orbits. In conclusion, this paper presents a solution to the problem of determining the time a satellite stays in the shadow of a celestial body, taking into account different models and complex trajectories. The outcomes from the solution are crucial for precise and effective calculations of orbits and serve as a useful tool for space mission planning.

# 1. Methods

## 1.1. Universal model for calculating satellite eclipse duration (SEEM model definition)

The objective of this section is to establish a universal model applicable to the determination of the duration of satellite eclipses in a conic model of shadows. The presented solution is universal for both elliptical and hyperbolic orbits and is a special solution for the parabolic case. Additionally, the method is suitable for solving both planar and spatial eclipse problems.

The model is built under the following assumptions:

1. Satellites revolve around the central body without perturbation (i.e., the trajectory of the satellite is a Keplerian orbit);

2. The celestial body and the Sun are assumed to have a perfectly spherical shape.
3. Aberration of light is not considered;
4. Atmospheric refraction is neglected;
5. Relativistic effects are neglected;
6. The motion of the Sun during a satellite orbit is not considered.

At first, it is required to describe the equations for finding parameters that in full describe the position of the Sun in the orbital plane and relative to it. One specifies such a coordinate system $X_\gamma, Y, Z$ in which the $X_\gamma, Y$ plane coincides with the plane of the planet's orbit[2] (Fig. 2). In this system, the position of the Sun (*S* in Fig. 2) is accurately determined by an angle $\lambda_\odot^*$ calculated counter clockwise in Fig. 2, assuming that nutation and precession of the orbital plane can be neglected. Let us also set a point on the line joining the centers of masses of the Sun and the planet, but is opposite to the Sun (*S** in Fig. 2). Then, the position of this point can be set with $\lambda_\odot = \pi - \lambda^*_\odot$. Then, one can find the position of $S^*$ in the orbital frame. To do this, the the angles $\varpi_\odot$ and $\beta_\odot$ should be obtained from the geometry of the diagram in Fig. 2

The angle between Sun-planet line and satellite orbit plane is defined as follows, solving the spherical triangle *BS*D* in Fig. 2 (side *S*D*):

(2) $$\sin\beta_\odot = -\sin\left(\lambda_\odot - \Omega_1\right)\sin i_1 .$$

The $\varpi_\odot$ can be defined as follows (see Fig. 2):

(3) $$\varpi_\odot = \omega_1 + \psi_\odot ,$$

where $\tan\psi_\odot = \tan\left(\lambda_\odot - \Omega_1\right)\cos i_1$. Note that $\varpi_\odot = \begin{cases} \varpi_\odot & |\omega_1 + \psi_\odot| \le \pi \\ 2\pi - \varpi_\odot & |\omega_1 + \psi_\odot| > \pi \end{cases}$.

The main assumption made in this work is that the orbital plane intersects the cone by angle $\beta_\odot$ to its axis and then shadows leaving the footprint on the orbital plane that can be described by a second-order curve. The proof that the shadow of a planet in the orbital plane can be described as an ellipse (hereafter referred to as a shadow ellipse) is provided in **Appendix A**. Fig. 3 shows the general scheme of satellite occultation by a spherical celestial body under the above assumptions. The interaction of the spherical body with the inclined orbital plane circular cone of the penumbra leaves a footprint of shadow as a part of the ellipse (darkened in the picture). The consideration of shadows as whole ellipses with both occulted and sun side parts provides a general case for all orbits of satellites with no regard to their shape. The only condition is that the orbit of the satellite should remain stable during at least one satellite revolution around the central body.

[2] Note that basic transformations of Keplerian elements of satellite orbit from any other frame to those specified above can be made from simple geometric relations provided in many classical works on celestial mechanics and astrodynamics; for example, see [30,33]. The position of the Sun in a specified frame can be calculated using ephemerides, e.g., DE441 (https://naif.jpl.nasa.gov/pub/naif/generic_kernels/spk/planets/ (Accessed Jul. 7, 2023)).

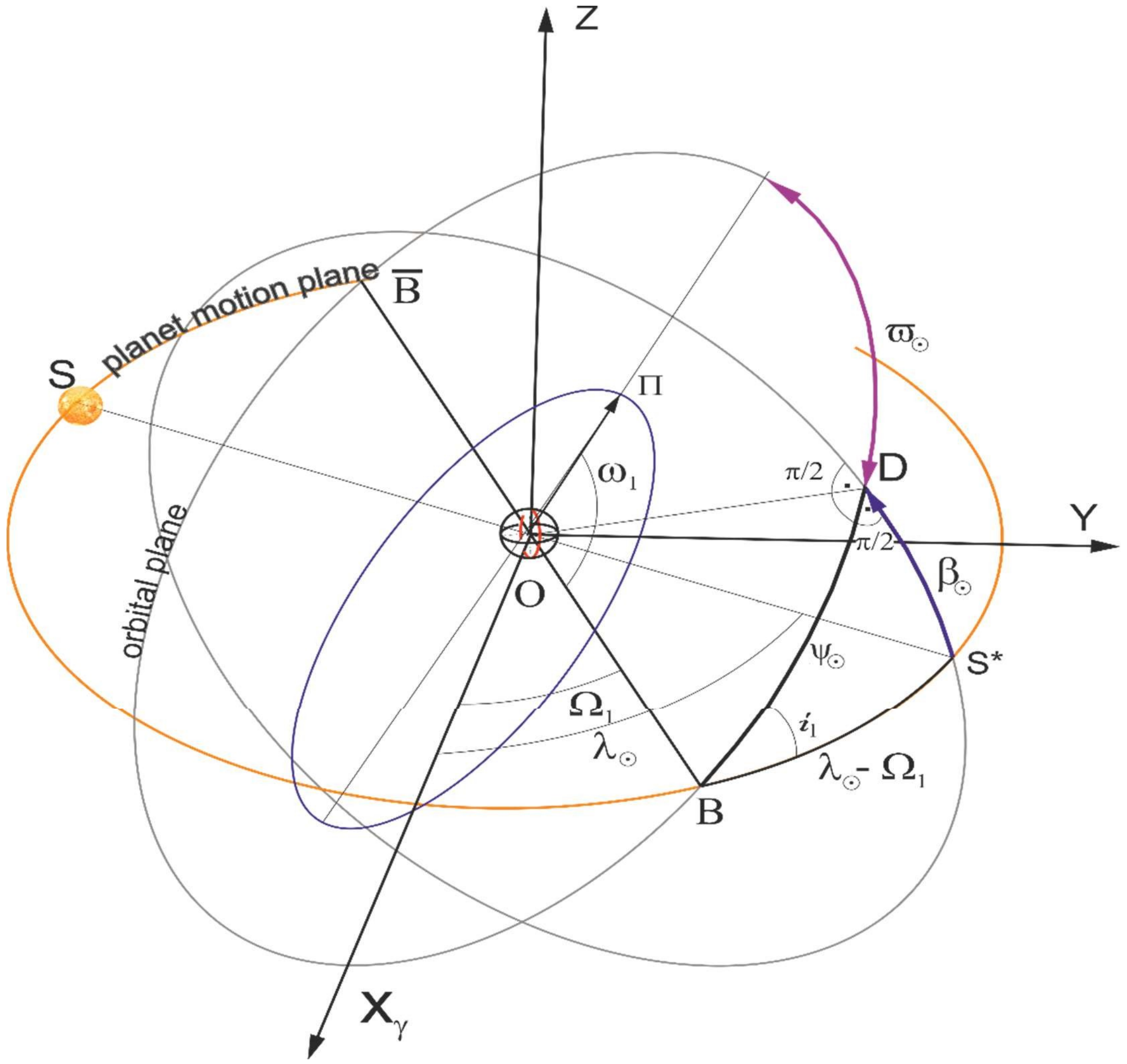

Figure 2. Scheme for deriving the required parameters for the Sun's orientation in a planet-centric frame

The base geometry for calculating the parameters of the satellite's orbit shadowing in the case of using the conical model and considering the combined case, i.e., for the satellite staying within the penumbra cone, is shown in Fig. 4, *a* and *b*. Since the penumbra cone includes a whole umbra region, the main problem is to provide the analytical solution to the penumbra case. When applicable, the solution for the umbra cone is also given.

Graphically, from the geometric construction in Fig. 4, the main parameters of the shadow ellipse can be obtained as functions of the Sun's position out of the orbital plane $\delta$ : $a_2 = \frac{R}{2}\left(\frac{\sin(\beta_\odot + \alpha_p) + \sin(\beta_\odot - \alpha_p)}{\sin(\beta_\odot + \alpha_p)\sin(\beta_\odot - \alpha_p)}\right)$, is the semimajor axis under the condition of nonsingularity solutions $|\beta_\odot| > |\alpha_p|$; $e_2 = \frac{\cos\beta_\odot}{\cos\alpha_p}$ is the eccentricity of the ellipse. The auxiliary parameter (semiminor axis of an ellipse) $b_2 = a_2\sqrt{1-e^2_2}$ is also considered. Note that in the case of $0 < |\beta_\odot| < |\alpha_p|$ and $|\beta_\odot| = |\alpha_p|$, the shadow is quasiparabolic. However, to simplify the solution and avoid singularities, a separate solution for these cases is provided.

Additionally, it can be shown that the parameters for the shadow ellipse can be simplified for the cylindrical shadow model $a_2 = \frac{R}{\sin\beta_\odot}$, $|\beta_\odot| > 0$, $e_2 = \cos\beta_\odot$, $b_2 = R$.

The problem of determining the part of the satellite orbit occulted by the celestial body is then reduced to determine the intersection points of both the satellite orbit ellipse in eq. (4) and the shadow ellipse (5). The canonical equations of the ellipses in the coordinate system centred on the geometric centre of the satellite orbit ellipse can be written as:

(4) $$\frac{\xi^2}{a_1^2} + \frac{\eta^2}{b_1^2} = 1,$$

(5) $$\frac{((\xi - \xi_0)\cos\varpi_\odot + (\eta - \eta_0)\sin\varpi_\odot)^2}{a_2^2} + \frac{(-(\xi - \xi_0)\sin\varpi_\odot + (\eta - \eta_0)\cos\varpi_\odot)^2}{b_2^2} = 1$$

Let us write the parametric equation for the ellipse ($\mathcal{E}_2$ in Fig. 4*a*) in coordinates $\xi\eta$ centred in *O'*:

(6) $$\begin{cases} \xi = \xi_0 + a_2\cos\theta\cos\varpi_\odot - b_2\sin\theta\sin\varpi_\odot \\ \eta = \eta_0 + a_2\cos\theta\sin\varpi_\odot + b_2\sin\theta\cos\varpi_\odot \end{cases},$$

where from the geometry presented in Fig. 4*b* $\xi_0 = a_1 e_1 + \|\overline{O'F}\|\cos\varpi_\odot$, $\eta_0 = \|\overline{O'F}\|\sin\varpi_\odot$, $\|\overline{O'F}\| = a_2 - R/\sin(\beta_\odot + \alpha_p)$.

Let us convert equation (6) considering transformation $\cos\theta = \frac{1-\tan^2\frac{\theta}{2}}{1+\tan^2\frac{\theta}{2}}$, $\sin\theta = \frac{2\tan\frac{\theta}{2}}{1+\tan^2\frac{\theta}{2}}$:

(7) $$\begin{cases} \xi = \dfrac{1}{1+\tan^2\dfrac{\theta}{2}}\left[\left(-a_2\cos\varpi_\odot+\xi_0\right)\tan^2\dfrac{\theta}{2}-2b_2\sin\varpi_\odot\tan\dfrac{\theta}{2}+a_2\cos\varpi_\odot+\xi_0\right] \\ \eta = \dfrac{1}{1+\tan^2\dfrac{\theta}{2}}\left[\left(-a_2\sin\varpi_\odot+\eta_0\right)\tan^2\dfrac{\theta}{2}+2b_2\cos\varpi_\odot\tan\dfrac{\theta}{2}+a_2\sin\varpi_\odot+\eta_0\right] \end{cases},$$

$\theta\in[0;2\pi)$, $\varpi_\odot\in[0;2\pi)$

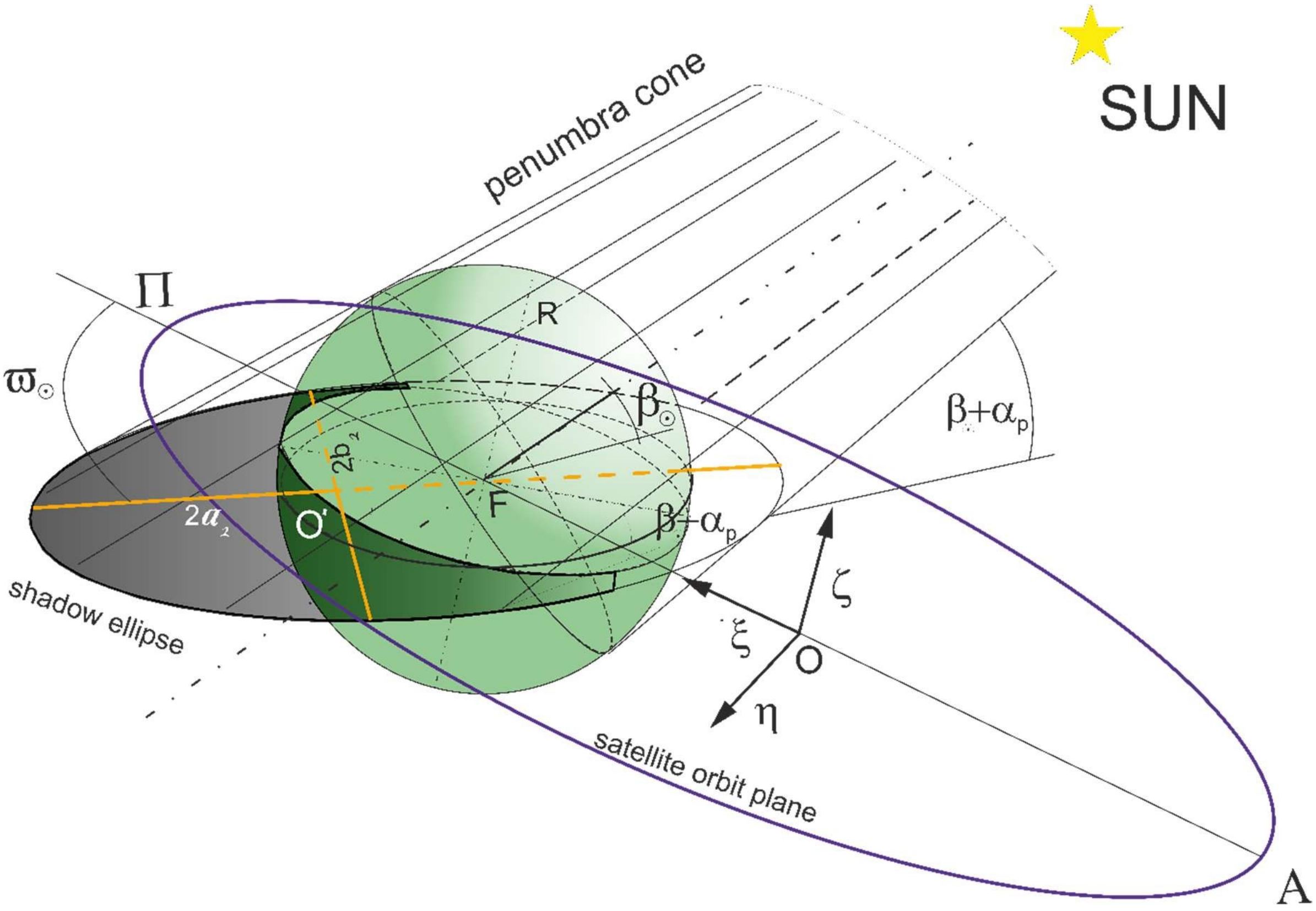

Figure. 3 General case for occultation, it is shown that the cone of the penumbra being cut by the orbital plane leaves a footprint on it, the elliptical part of which is obscured by the body, while the other part is on the Sun side. $\Pi, A$ designate the periapsis and apoapsis of the satellite orbit.

Substituting the obtained expressions for coordinates of points of intersection (7) in eq. (4) and denoting $\tilde{x}=\tan\frac{\theta}{2}$, the following equation of the 4th degree is obtained:

(8) $$s_0\tilde{x}^4+s_1\tilde{x}^3+s_2\tilde{x}^2+s_3\tilde{x}+s_4=0,$$

The obtained Eq. (8) is similar to those studied in [1,8,9,24]. However, unlike previous approaches, this one is written relative to the parameter $\tilde{x}=\tan\frac{\theta}{2}$, where $\theta$ is the angle between the radius of the shadow ellipse calculated counterclockwise from the direction opposite to the Sun. Ergo, it makes it possible to further distinguish the roots of the 4th degree equation that are chosen as entry and exit points from the penumbra region.

It should also be noted that numerical as well as analytical and numerical studies have been carried out in [25–27], where the authors have provided fast and reliable methods for solving the quartic equation.

The coefficients in (8) for hyperbolic and elliptical satellite orbits (hyperbolic coefficients obtained in the same way as described for elliptic orbits) in simplified form can be written as follows:

$$
\begin{aligned}
s_0 &= a_1^2 a_2^2 k_2 \tilde{\varsigma}_1 - 2a_1^2 a_2 k_2 y_0 \tilde{\varsigma}_1 - a_1^2 b_1^2 + a_1^2 y_0^2 \tilde{\varsigma}_1 + a_2^2 b_1^2 k_1^2 - 2a_2 b_1^2 k_1 x_0 + b_1^2 x_0^2 \\
s_1 &= -4a_1^2 a_2 b_2 k_1 k_2 \tilde{\varsigma}_1 + 4a_1^2 b_1 k_1 y_0 \tilde{\varsigma}_1 + 4a_2 b_1^2 b_2 k_1 k_2 - 4b_1^2 b_2 k_2 x_0 \\
s_2 &= -2a_1^2 a_2^2 k_2^2 \tilde{\varsigma}_1 - 2a_1^2 b_1^2 + 2a_1^2 y_0^2 \tilde{\varsigma}_1 - 2a_2^2 b_1^2 k_1^2 + 2b_1^2 x_0^2 + 4a_1^2 b_2^2 k_1^2 \tilde{\varsigma}_1 + 4b_1^2 b_2^2 k_2^2 , \\
s_3 &= 4a_1^2 a_2 b_2 k_1 k_2 \tilde{\varsigma}_1 + 4a_1^2 b_2 k_1 y_0 \tilde{\varsigma}_1 - 4a_2 b_1^2 b_2 k_1 k_2 - 4b_1^2 b_2 k_2 x_0 \\
s_4 &= a_1^2 a_2^2 k_2^2 \tilde{\varsigma}_1 + 2a_1^2 a_2 k_2 y_0 \tilde{\varsigma}_1 - a_1^2 b_1^2 + a_1^2 y_0^2 \tilde{\varsigma}_1 + a_2^2 b_1^2 k_1^2 + 2a_2 b_1^2 k_1 x_0 + b_1^2 x_0^2
\end{aligned}
\tag{9}
$$

$$\tilde{\varsigma}_1 = sign(1 - e_1),\ e_1 \neq 1$$

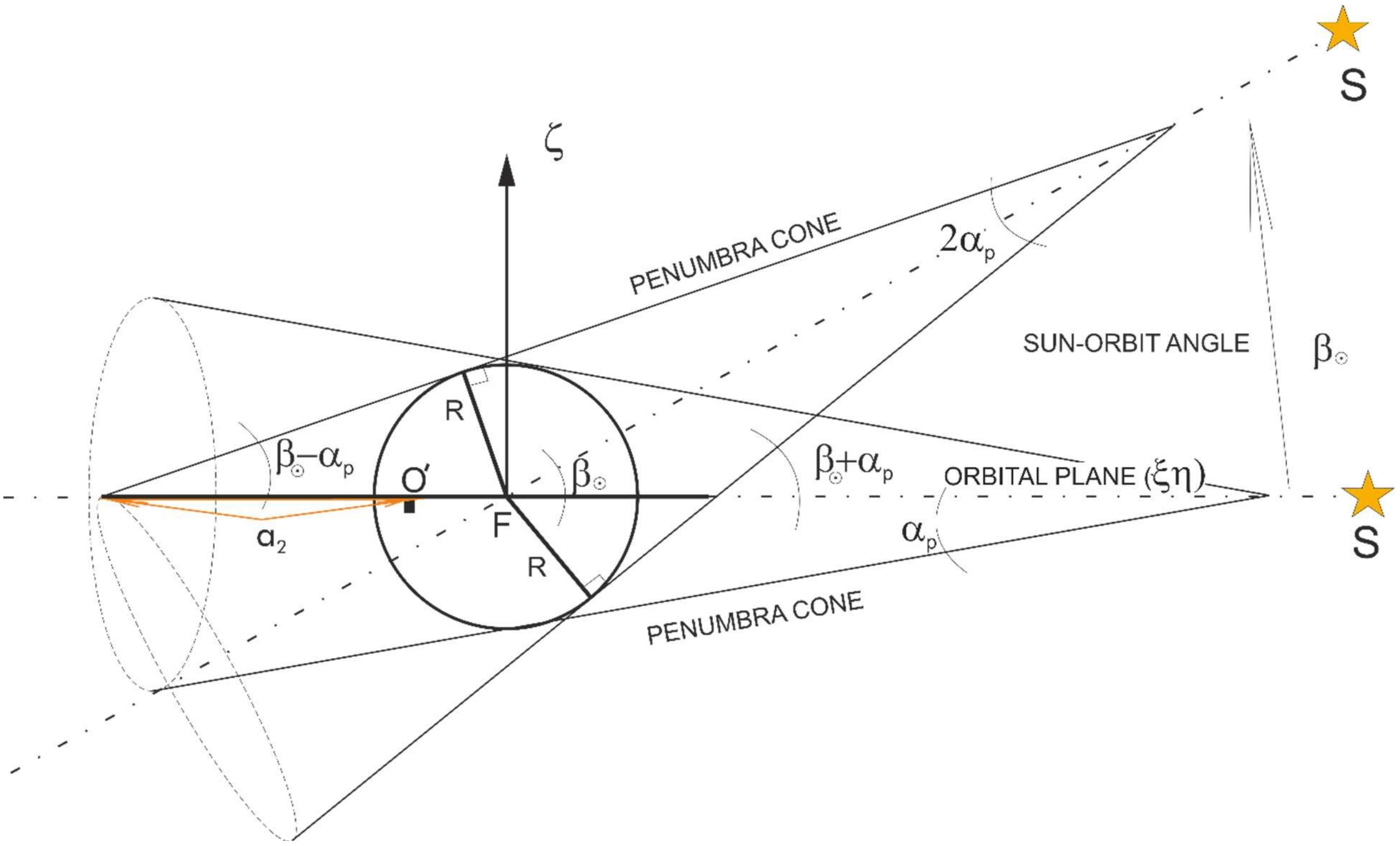

(a)

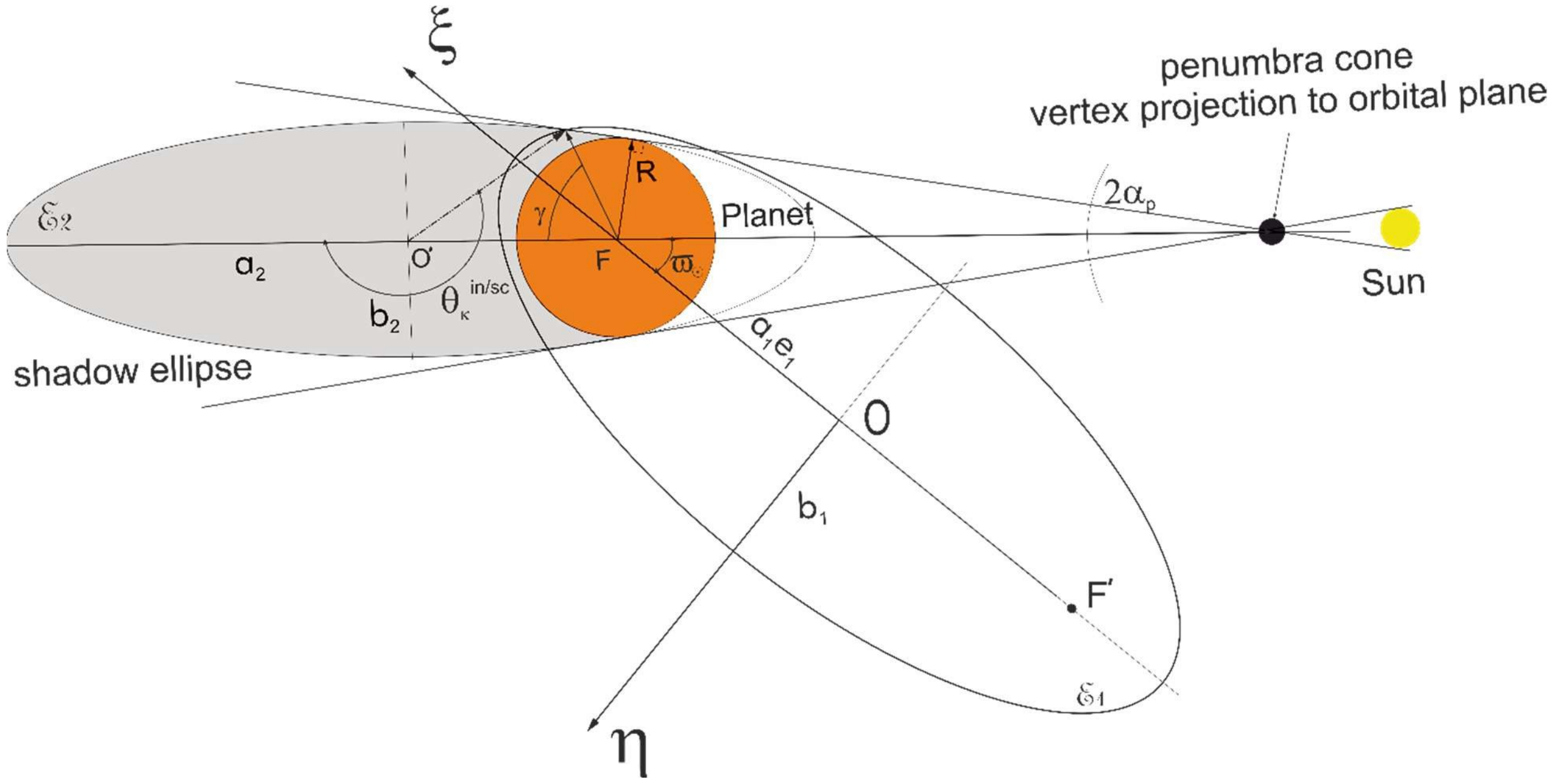

(b)

Figure. 4 (a) Geometry of the calculation parameters of the curve that appears in the satellite orbital plane after intersection with the penumbra cone. The direction of view is perpendicular to the plane in which the penumbra cone rotation axis is placed. (b) The shadow curve in the satellite orbit plane as viewed from the $\zeta$ axis. FF' is the foci of the orbital ellipse

The required functions for coefficients (9) were already established earlier in the text, but for the reader convenience, they are repeated here:

$$a_2 = \frac{R}{2}\left(\frac{\sin(\beta_\odot + \alpha_p) + \sin(\beta_\odot - \alpha_p)}{\sin(\beta_\odot + \alpha_p)\sin(\beta_\odot - \alpha_p)}\right),\ e_2 = \frac{\cos\beta_\odot}{\cos\alpha_p},\ b_2 = a_2\sqrt{1-e^2{}_2}, k_1 = \cos\varpi_\odot,\ k_2 = \sin\varpi_\odot,$$

$$\varpi_\odot = \omega_1 + \psi_\odot, \sin\beta_\odot = -\sin(\lambda_\odot - \Omega_1)\sin i_1, \cos\beta_\odot = \sqrt{1-\sin^2(\lambda_\odot - \Omega_1)\sin^2 i_1},$$

$$\xi_0 = a_1 e_1 + (a_2 - R/\sin(\beta_\odot + \alpha_p))\cos\varpi_\odot, \eta_0 = (a_2 - R/\sin(\beta_\odot + \alpha_p))\sin\varpi_\odot, b_1 = a_1\sqrt{\left|1-e_1^2\right|}$$

Thus, the solutions for eq. (8)can be written as follows:

(10) $$\tilde{x}_\kappa = F_\kappa\left(\{a_1, e_1, i_1, \Omega_1, \omega_1\}, \lambda_\odot\right), \kappa = \overline{1,4}$$

Solutions (10) correspond to shadow regions, where $|\beta_\odot| > |\alpha_p|$. In the case of $0 < |\beta_\odot| < |\alpha_p|$, solutions can be found by solving the second-degree polynomial function two times to find the intersection points between the orbital curve of the spacecraft trajectory and the pair of direct lines that represent the shadow in the orbital plane. The required polynomial function and corresponding coefficients are given below as well as the parametric equation required for calculating intersection points between discussed curves (in this form $\tilde{x} = \theta$):

(11) $$s_2\tilde{x}^2 + s_3\tilde{x} + s_4 = 0$$

where

$$s_2 = a_1^2 a_2^2 k_2^2 \tilde{\varsigma}_1 - 2a_1^2 a_2 b_2 k_1 k_2 \tilde{\varsigma}_2 - a_1^2 b_2^2 k_1^2 \tilde{\varsigma}_1 + a_2^2 b_1^2 k_1^2 + 2a_2 b_1^2 b_2 k_1 k_2 \tilde{\varsigma}_2 \tilde{\varsigma}_1 + b_1^2 b_2^2 k_2^2$$

$$s_3 = -2a_1^2 b_2 k_1 y_0 \tilde{\varsigma}_1 + 2a_1^2 b_2 k_1 y_0 \tilde{\varsigma}_2 - 2a_2 b_1^2 k_1 x_0 - 2b_1^2 b_2 k_2 x_0 \tilde{\varsigma}_2 \tilde{\varsigma}_1 \qquad , \tilde{\varsigma}_2 = \begin{cases} 1, & \kappa = 1 \\ -1, & \kappa = 2 \end{cases},$$

$$s_4 = -a_1^2 b_1^2 + a_1^2 y_0^2 \tilde{\varsigma}_1 + b_1^2 x_0^2$$

Parametric equations for direct lines in the conferred case are:

$$\begin{cases} \xi = \xi_0 - \tilde{a}_2 \theta \cos\varpi_\odot \pm_{\tilde{\kappa}} \tilde{b}_2 \theta \sin\varpi_\odot \\ \eta = \eta_0 - \tilde{a}_2 \theta \sin\varpi_\odot \mp_{\tilde{\kappa}} \tilde{b}_2 \theta \cos\varpi_\odot \end{cases}, \quad \tilde{\kappa} = 1, 2, \tag{12}$$

where $\xi_0 = a_1 e_1 + \tilde{a}_2 \cos\varpi_\odot$, $\eta_0 = \tilde{a}_2 \sin\varpi_\odot$, $\tilde{a}_2 = R / \tan\alpha_p$, $\tilde{b}_2 = R$

The combined solution for solving equation (11) two times with different coefficients $s_3, s_4$ distinguished by $\tilde{\varsigma}_2$ can be written in the form of (10).

Note that coefficients $s_0, s_1, s_2, s_3,$, and $s_4$ for the cylindrical model of the body's shadow can be obtained by assuming $\eta_0 = 0$ and $\xi_0 = a_1 e_1$. Thus, equation (8) is universal for both elliptic and hyperbolic orbits of satellites as well as for both conical and cylindrical models of shadows.

The solution to parabolic satellite orbits ($e_1 = 1$) can be derived using the direct substitution of $\xi = \eta^2 / 2p_1$ in equation (5) and obtaining the 4th degree equation of $\eta$ :

$$\tilde{s}_0 \eta^4 + \tilde{s}_1 \eta^3 + \tilde{s}_2 \eta^2 + \tilde{s}_3 \eta + \tilde{s}_4 = 0 \tag{13}$$

where coefficients can be denoted as follows:

$$\begin{aligned} \tilde{s}_0 &= a_2^2 p_1^2 k_2^2 + b_2^2 p_1^2 k_1^2 \\ \tilde{s}_1 &= 4a_2^2 p_1 k_1 k_2 + 4b_2^2 p_1 k_1 k_2 \\ \tilde{s}_2 &= a_2^2 k_1^2 - a_2^2 k_1 k_2 p_1 \eta_0 + a_2^2 k_2^2 p_1 \xi_0 + b_2^2 k_1 k_2 p_1 \eta_0 + b_2^2 k_2^2 \\ \tilde{s}_3 &= -2a_2^2 k_1^2 \eta_0 + 2a_2^2 k_1 k_2 \xi_0 - 2b_2^2 k_1 k_2 \xi_0 - 2b_2^2 k_2^2 \eta_0 \\ \tilde{s}_4 &= -a_2^2 b_2^2 + a_2^2 k_1^2 \eta_0^2 - 2a_2^2 k_1 k_2 \xi_0 \eta_0 + a_2^2 k_2^2 \xi_0^2 + b_2^2 k_1^2 \xi_0^2 + 2b_2^2 k_1 k_2 \xi_0 \eta_0 + b_2^2 k_2^2 \eta_0^2 \end{aligned} \tag{14}$$

The analytical solution of the above equation is simply derived using any method for solving the 4th-degree algebraic equation.

Once the solutions to the above equations (8, 11) are obtained, they can be placed in the original parametric equation of the shadow curve (ellipse for (8), lines for (11)) to obtain the coordinates of the intersection points. For parabolic orbits, the solutions in true anomalies are easily found as:

$${}^{p}f_{1,\kappa}^{\,in/sc} = \arctan\frac{\eta_\kappa}{\xi_\kappa(\eta_\kappa)}, \quad \kappa = 1,...,4 \tag{15}$$

The true anomaly of each intersection point of the satellite (elliptic or hyperbolic) orbit with the penumbra cone can be determined by the following equation (for distinguishing the solution for the penumbra from the umbra, the left upper script "*p*" is used):

(16) $$ {}^{p}f_{1,\kappa}^{\ in/sc} = \begin{cases} \arctan\dfrac{(\eta_0 - a_2k_2)\tilde{x}_\kappa^{\ 2} + 2b_2k_1\tilde{x}_\kappa + a_2k_2 + \eta_0}{(\xi_0 - a_1e_1 - a_2k_1)\tilde{x}_\kappa^{\ 2} - 2b_2k_1\tilde{x}_\kappa + a_2k_1 - a_1e_1 + \xi_0}, & |\beta_\odot| < |\alpha_p| \\ \arctan\dfrac{\eta_0 - (a_2k_2 \mp_\kappa b_2k_1)\tilde{x}_\kappa}{\xi_0 - (a_2k_1 \pm_\kappa b_2k_2)\tilde{x}_\kappa}, & 0 < |\beta_\odot| < |\alpha_p| \end{cases}, \quad \kappa = 1,...,4 $$

Out of the 4 possible solutions, those that are the closest to the desired value of $\varpi_\odot$, i.e., satisfying constraint $|\gamma| = \left|\arccos\dfrac{\mathbf{r}_1({}^{p}f_{1,\kappa}^{\ in/sc})\bullet\mathbf{r}^*_\odot}{r_1({}^{p}f_{1,\kappa}^{\ in/sc})r^*_\odot}\right| \le \dfrac{\pi}{2}$, $\kappa = 1,...,4$, , should be selected. To check the spurious roots, the above-mentioned advantage of finding the solution in $\theta$ can be used:

(17) $$ \begin{cases} \kappa = \overline{3,4}, & \varpi_\odot \in \left(\dfrac{3\pi}{2}; 2\pi\right] \\ \kappa = \overline{2,3}, & \varpi_\odot \in \left(\dfrac{\pi}{2}; \dfrac{3\pi}{2}\right] \\ \kappa = \overline{1,2}, & \varpi_\odot \in \left(0; \dfrac{\pi}{2}\right] \end{cases} $$

The final true anomaly after all of the above calculations can be written in universal solution form ($\tilde{F}_{+/-}$):

(18) $$ {}^{p}f_{1,+/-} = \tilde{F}_{+/-}\left(\{a_1, e_1, i_1, \Omega_1, \omega_1\}, \lambda_\odot\right) $$

After determining the true anomaly of the satellite at which it can be on the penumbra border, the eccentric anomaly can be calculated as follows [28]:

(19) $$ {}^{p}E_{1,+/-} = 2\arctan\left[\sqrt{\frac{1-e_1}{1+e_1}}\tan\frac{{}^{p}f_{1,+/-}}{2}\right], \ e < 1 $$

To calculate the time of flight along the hyperbolic orbit, the hyperbolic anomaly should be obtained:

(20) $$ {}^{p}H_{1,+/-} = 2\,\mathrm{arctan\,h}\left[\sqrt{\frac{e_1-1}{e_1+1}}\tan\frac{{}^{p}f_{1,+/-}}{2}\right], \ e > 1 $$

The time of flight, then, might be determined as [28,29]:

(21) $$ \Delta t = \begin{cases} \sqrt{\dfrac{a_1^3}{\mu_{pl}}}\left(e_1\sin{}^{p}E_{1,+} - {}^{p}E_{1,+} - e_1\sin{}^{p}E_{1,-} + {}^{p}E_{1,-}\right), & e_1 < 1 \\ \dfrac{p_1^{3/2}}{\sqrt{\mu_{pl}}}\left(\tan\dfrac{{}^{p}f_{1,+}}{2} + \dfrac{1}{3}\tan^3\dfrac{{}^{p}f_{1,+}}{2} - \tan\dfrac{{}^{p}f_{1,-}}{2} - \dfrac{1}{3}\tan^3\dfrac{{}^{p}f_{1,-}}{2}\right), & e_1 = 1 \\ \sqrt{\dfrac{a_1^3}{\mu_{pl}}}\left(e_1\sinh{}^{p}H_{1,+} - {}^{p}H_{1,+} - e_1\sinh{}^{p}H_{1,-} + {}^{p}H_{1,-}\right), & e_1 > 1 \end{cases} $$

Thus, the time of satellite flight in the shadow of a planet (combined, i.e., in the penumbra, which includes the umbra region) depending on its orbital parameters has been obtained.

**Remark 1.** The solutions presented above are sufficient for finding the shadow time for the penumbra and umbra regions (i.e., in part of the shadows and complete shadows). To find the solution for the time of flight of the satellite in the umbra region, it is necessary to repeat the steps presented above, replacing $\alpha_p$ by $\alpha_u$ in $a_2 = \dfrac{R}{\sin(\beta_\odot - \alpha_u)}$, and in $e_2 = \dfrac{\cos\beta_\odot}{\cos\alpha_u}$. Thus, the value of $f_{1,\kappa}^{in/sc}$ closest to the opposite direction of $\varpi_\odot$, unlike the previous case, should be chosen.

**Remark 2.** It is easy to identify the parameters of satellite ellipse $(e_1, p_1)$ for given $\beta_\odot, \varpi_\odot$ at which the solution to Eq. (18) exists. To do this, the following condition should be:

$$\|\overline{FA_2}\| \geq r_1(f_1 = \tilde{\varpi}_\odot), \tag{22}$$

where $\|\overline{FA_2}\|$ is the distance from the planet centre to the apoapsis point of the shadow ellipse.

Let us rewrite the equation as follows

$$\frac{R}{\left|\sin(\beta_\odot - \alpha_p)\right|} \geq \frac{p_1}{1 + e_1\cos\tilde{\varpi}_\odot}, \tag{23}$$

Then, the above inequality can be rewritten

$$\left\{\begin{matrix}\left(R/p_1 - \sin(\beta_\odot - \alpha_p)\right)\cos\beta_\odot + \\ e_1 R/p_1\left(\cos(\lambda_\odot - \Omega_1)\cos\omega_1 + \sin(\lambda_\odot - \Omega_1)\sin\omega_1\cos i_1\right)\end{matrix}\right\} \geq 0, \tag{24}$$

In the above inequality the replacement of $\cos\varpi_\odot$ is used

$$\cos\varpi_\odot = \left(\cos(\lambda_\odot - \Omega_1)\cos\omega_1 + \sin(\lambda_\odot - \Omega_1)\sin\omega_1\cos i_1\right)/\cos\beta_\odot.$$

If the above criteria are met, then, given the Keplerian elements of a satellite's orbit, a shadow zone will exist and it will be possible to find a solution to the shadow equation.

**Remark 3.** The solution for the near plane problem, i.e., when $|\beta_\odot| \leq |\alpha_p|$, can be further simplified for an elliptical orbit with eccentricity $e_1 < 1$ and obtained as a combination of the Taylor series form. This case is considered in **Appendix B.**

## 1.2 Expansion of the model with an adaptation for oblateness

The introduction, already listed the studies that made a point to establish models for predicting satellite eclipses in the case of oblateness. One of the most widely cited is that of Adhya et al [12], where the authors presented a common case solution that has been the basis for other established models [15-17]. This paper proposes a different approach to solving the problem in the case of a satellite orbiting a non-spherical body. It is based on the model presented in the previous section. For this purpose, it is necessary to obtain a parameter of the shadow ellipse (i.e. $a_2, e_2$) in a new form with the strict condition that such determined parameters should be a function of the satellite's orbital parameters and the Sun's longitude. The following assumptions should be made for the construction of the model:

1. The sunlight rays are parallel to the direction of the sun-planet line.
2. The shape of the planet is described by a spheroid (i.e. a two-axis ellipsoid).
3. The Sun is assumed to have a perfectly spherical shape.
4. Atmospheric refraction is neglected.
5. Relativistic effects are not taken into account.
6. The motion of the Sun during satellite orbit is not considered.

Notice that the parameters of the shadow ellipse $a_2, e_2$ in this section are denoted as $a^o{}_2, e^o{}_2$ where "o" stands for oblateness.

Let us consider a sphere of radius $R_{eq}$ (Fig. 5), which occludes the sunlight rays spreading parallel to the sun-planet line. Let a spheroid with the parameters $a_{pl}$ , $b_{pl}$ , and $e_{pl} = \sqrt{1 - \frac{b_{pl}^2}{a_{pl}^2}}$, which are the semimajor and semiminor axes and the eccentricity, be inscribed in the sphere. As an example of the mentioned parameters, the reader is referred to those for the Earth spheroid according to (IERS)[3] [25, p. 75]: $a_{pl} = 6{,}378.136\,\text{km}$ , $b_{pl} = 6{,}356.751\,\text{km}$ and $e_{pl} = 0.08182$ .

Let the shadow cylinder, sphere describing the shape of planet at first approximation and as well as spheroid to be intersected by a plane that contains a straight line connecting the centres of mass of the Sun and the planet (Fig. 5). The angle between the equatorial plane of the spheroid and the orbital plane of the satellite is denoted by $\tilde{\varphi} \in [0;\pi)$. The angle $\tilde{\alpha} = \frac{\pi}{2} - \delta$ denotes the position of the line *FL* connecting the centre of mass of the planet and the point of tangency of the solar ray (*L*) and the sphere $R_{eq}$ .

[3] https://iers-conventions.obspm.fr/ (Accessed 25 June, 2023).

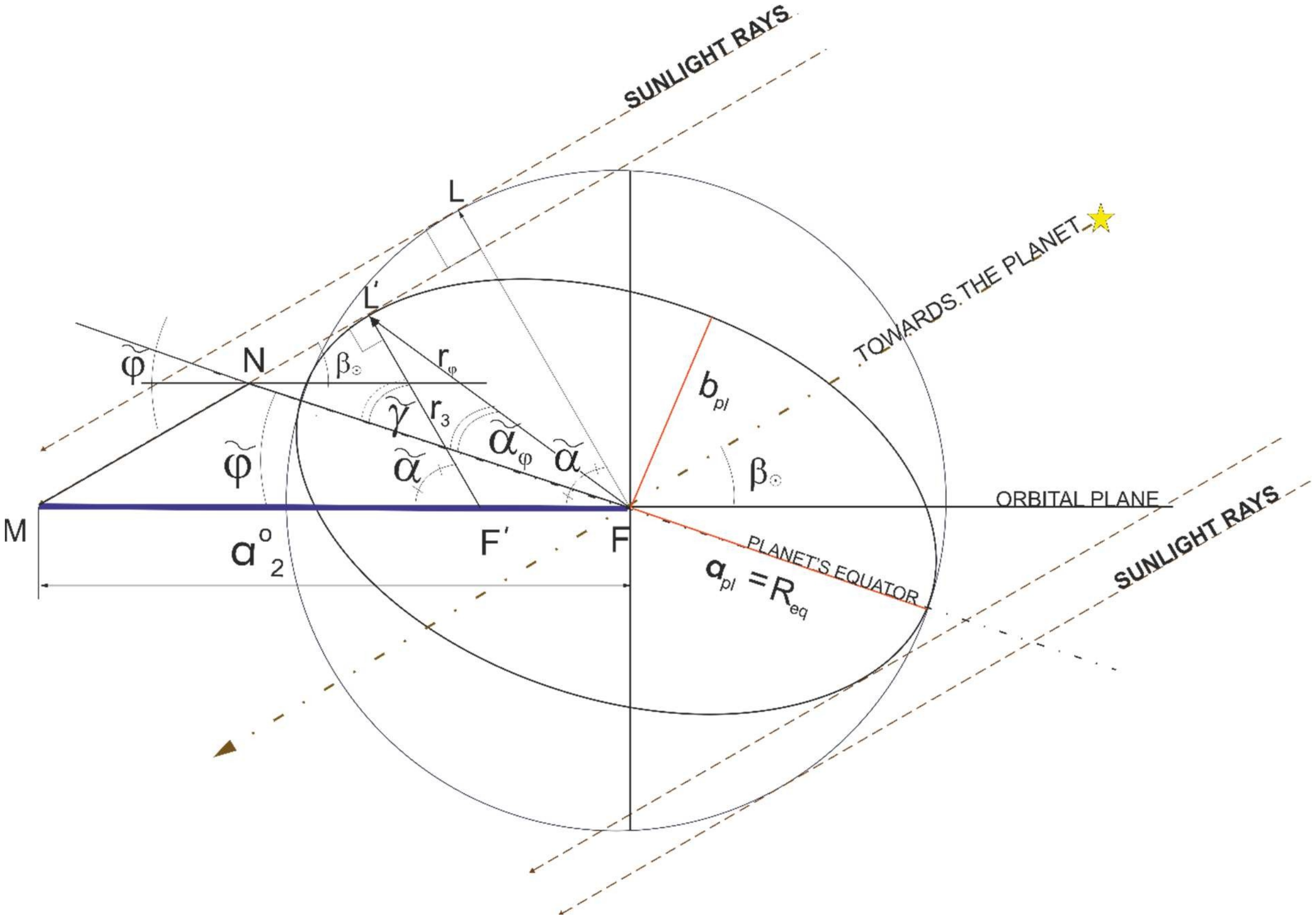

Figure. 5. Scheme for calculating the parameters of the shadow ellipse in the case of a spheroid shape of a planet (figures show the intersection of planet's spheroid by plane that is containing Sun-planet line and that is orthogonal to orbital plane).

Furthermore, knowing the point *L*, one can easily obtain the tangent point of the ray to the spheroid (*L'*) and draw through this point the line *F'L'*, which intersects the plane of the satellite orbit at an angle $\tilde{\alpha}$ and the plane of the planet equator at an angle $\tilde{\gamma} = \frac{\pi}{2} - \left(\beta_{\odot} + \tilde{\varphi}\right)$.

Let us draw a straight line from the centre of mass of planet *F* to point *L'*, the length of which is denoted by $r_{\varphi}$. Let the angle between the plane of the planet equator and *FL'* be $\tilde{\alpha}_{\varphi}$. The relation that is well known in geodesy to establish the relationship between planetodetic coordinates and planetographic coordinates in this work is used to find the $\tilde{\alpha}_{\varphi}$ [30]:

$$\tan\tilde{\alpha}_{\varphi} = \sqrt{1-e_{pl}^2}\tan\tilde{\gamma} \quad (25)$$

After defining the necessary angle, it is possible to calculate the distance from the centre of the planet spheroid to the point of interaction between sunlight rays and the surface of the spheroid $r_{\varphi}$:

$$r_{\varphi} = \frac{b_{pl}}{\sqrt{1-e_{pl}^2\cos^2\tilde{\alpha}_{\varphi}}}, \quad (26)$$

Finally, the distance $r_3$ can be obtained by projecting both distances to the vertical line to the orbital plane:

(27)
$$r_3 = r_\varphi \frac{\sin\left(\tilde{\alpha}_\varphi + \tilde{\varphi}\right)}{\sin\tilde{\alpha}}.$$

The parameters of the shadow ellipse can be defined through the geometry of triangle *MFN* shown in Fig. 5.

(28)
$$e_2^0 = \cos\beta_\odot,$$

$$a^o{}_2 = \begin{cases} \dfrac{b_{pl}}{\sqrt{1-e_{pl}^2\cos^2\tilde{\alpha}_\varphi}}\left[\sin\left(\tilde{\alpha}_\varphi+\tilde{\varphi}\right)\tan\tilde{\alpha}+\cos\left(\tilde{\alpha}_\varphi+\tilde{\varphi}\right)-\sin\left(\tilde{\alpha}_\varphi+\tilde{\varphi}\right)\tan^{-1}\tilde{\alpha}\right], & 0<\left|\beta_\odot\right|<\dfrac{\pi}{2} \\ a_{pl}, & \left|\beta_\odot\right|=\dfrac{\pi}{2} \end{cases}.$$

The simplification of the resulting $a^o{}_2$ in expression (28) can be made by taking the first term of the Taylor series for $\sqrt{1-e_{pl}^2\cos^2\tilde{\alpha}_\varphi}$, since $\left|e_{pl}^2\cos^2\tilde{\alpha}_\varphi\right|<1$, because for most of the planets $e_{pl}\sim10^{-2}$.

(29)
$$a^o{}_2 = \begin{cases} b_{pl}\left(1+\dfrac{1}{2}e_{pl}^2\cos^2\left(\tilde{\alpha}_\varphi\right)\right)\left[\sin\left(\tilde{\alpha}_\varphi+\tilde{\varphi}\right)\tan\tilde{\alpha}+\cos\left(\tilde{\alpha}_\varphi+\tilde{\varphi}\right)-\sin\left(\tilde{\alpha}_\varphi+\tilde{\varphi}\right)\tan^{-1}\tilde{\alpha}\right], & 0<\left|\beta_\odot\right|<\dfrac{\pi}{2} \\ a_{pl}, & \left|\beta_\odot\right|=\dfrac{\pi}{2} \end{cases}.$$

Note that although the semimajor axis ($a^o{}_2$) of the shadow ellipse was defined, the eccentricity $e_2^0$ is the same as $e_2$ for the cylindric model of the shadow.

Having obtained the parameters of the shadow's ellipse, one can repeat the actions described in section 1.1 (replacing $a_2, e_2$ in eq. (9) coefficients *A, B, C, D,* and *E* with $a_2^o, e_2^0$) and calculate the time of satellite motion in the planet's shadow using eq. (21).

There is a special case for equatorial plane orbits, i.e., when $\tilde{\varphi}=0$ deg, then the above equation is simplified:

(30)
$$a^o{}_2 = \begin{cases} b_{pl}\left(1+\dfrac{1}{2}e_{pl}^2\cos^2\left(\tilde{\alpha}_\varphi\right)\right)\left[\sin\left(\tilde{\alpha}_\varphi\right)\tan\tilde{\alpha}+\cos\left(\tilde{\alpha}_\varphi\right)-\sin\left(\tilde{\alpha}_\varphi\right)\tan^{-1}\tilde{\alpha}\right], & 0<\left|\beta_\odot\right|<\dfrac{\pi}{2} \\ a_{pl}, & \left|\beta_\odot\right|=\dfrac{\pi}{2} \end{cases}$$

Note that the above statement of the problem allows us to obtain fairly simple expressions, which can then be used in section 1.1 to obtain the solution of equation (8) in a finite form. In the case of a conic shadow, all the above calculations remain valid, but the form of the obtained expressions becomes significantly more complicated.

### 1.3 Special case of simplification for elliptical orbits.

In section 1.1, the orbital plane intersects the penumbra cone, resulting in a shadow on the orbital plane that takes the form of an ellipse. The parameters of this ellipse can be determined by utilizing the Sun's longitude and the Keplerian elements of the satellite orbit. The purpose of this section is to derive a few cases where SEEM model can be simplified using the cylinder shadow model. To accomplish this, one should solve the problem of redefining coefficients *A, B, C, D,* and *E* once again. Consequently, the coefficients will be much simpler in their new form. It is important to note that this simplification is applicable only to the cylindrical shadow model. The main geometric constructions required in this section, which are utilized further, are depicted in Fig. 6.

The main assumptions for establishing a model presented in this section are the same as those described in section 1.1. Another one that should be added to those is the satellite orbit considered elliptical in this section. Additionally, the shadow creates a cylinder with a cross-diameter equal to twice the radius of the Earth.

As Fig. 6 shows, the following equation is satisfied at the intersection of the two ellipses:

(31) $$r_1^2 = r_2^2,$$

where $r_2 = \frac{b_2}{\sqrt{1 - e_2^2 \cos^2 \gamma}}$ is the distance from the centre to the point on the shadow ellipse.

Notice that the shadow equation (i.e., in the form of (8)) is described using the point of exit from the planet shadow since the resulting equation is the same if one starts from the entry point.

Equation (31) might be rewritten as:

(32) $$\left(1 + e_1 \cos f_1^{in/sc}\right)^2 = \chi\left(1 - e_2^2 \cos^2 \gamma_+\right),$$

where $\chi = \frac{a_1^2(1-e_1^2)^2}{a_2^2(1-e_2^2)}$ and $\gamma_+$ is the value of $\gamma$ at the intersection point of the satellite exit from the planet's shadow.

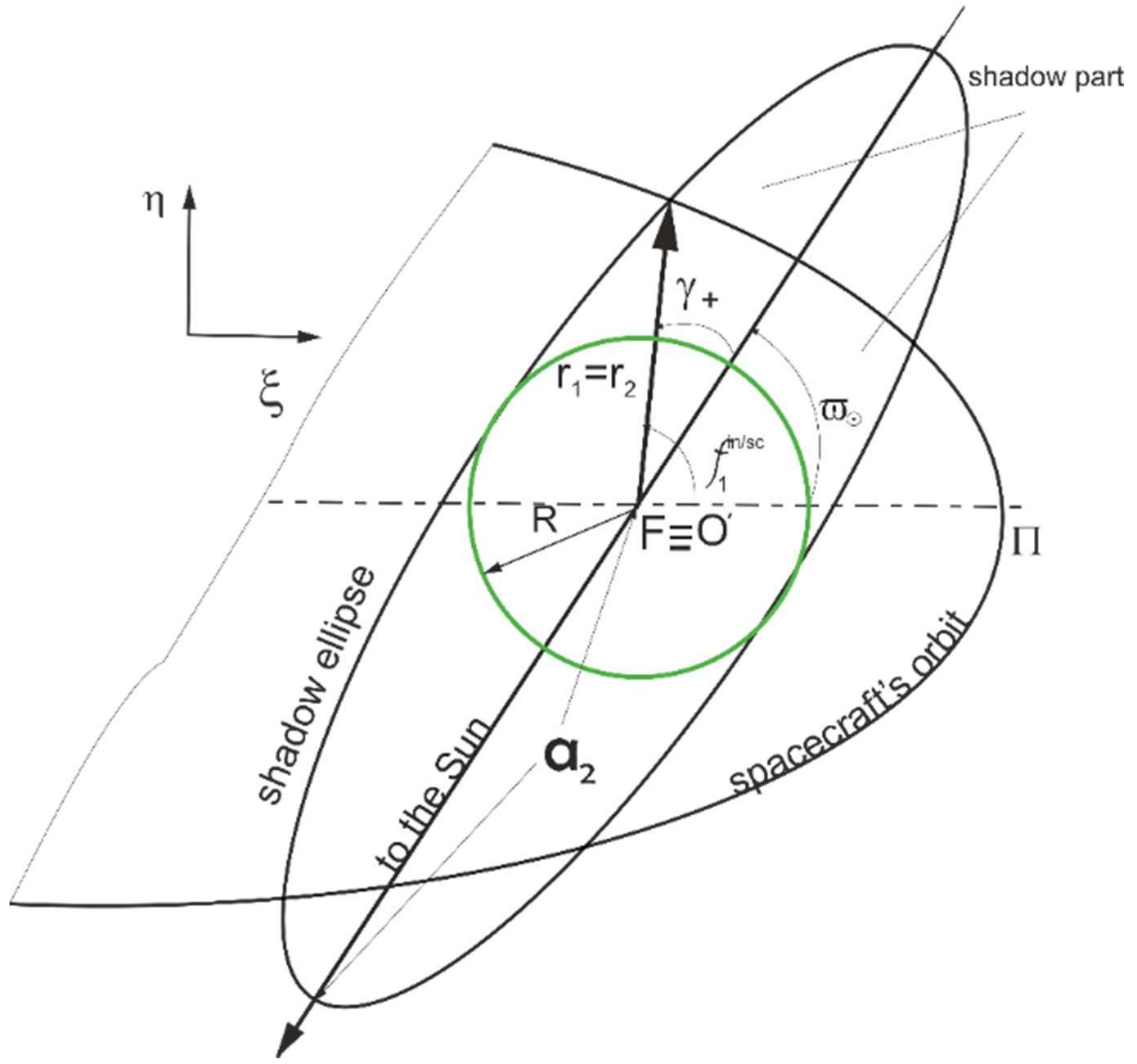

Figure 6. Geometry scheme for calculation. Π denotes periapsis. $\varpi_{\odot}$ is counted from the direction opposite to the direction of the Sun, counter clockwise.

After simplifying equation (32), considering that at the point of intersection $\gamma_{+} = f_1^{in/sc} - \varpi_{\odot}$ and using universal substitution, one obtains the 4th-order equation in the form of (8) written for the true anomaly:

(33) $$\tilde{A}\,\tilde{y}^4 + \tilde{B}\tilde{y}^3 + \tilde{C}\tilde{y}^2 + \tilde{D}\tilde{y} + \tilde{E} = 0\,,$$

where $\tilde{y} = \tan\frac{f_1^{in/sc}}{2}$ and the coefficients for this equation are found as follows:

(34) $$\begin{aligned}
\tilde{A} &= \frac{1}{2}\chi e_2^{\,2}\left(\cos 2\varpi_{\odot} + 1\right) - \chi + \left(1 - e_1\right)^2 \\
\tilde{B} &= -2\chi e_2^2 \sin 2\varpi_{\odot} \\
\tilde{C} &= -3\chi e_2^{\,2}\left(\cos 2\varpi_{\odot} - 1/3\right) + 2\left(1 - e_1^2 - \chi\right) \\
\tilde{D} &= 2\chi e_2^2 \sin 2\varpi_{\odot} \\
\tilde{E} &= \frac{1}{2}\chi e_2^{\,2}\left(\cos 2\varpi_{\odot} + 1\right) - \chi + \left(1 + e_1\right)^2
\end{aligned}$$

Solving this equation directly gives the true anomaly of intersection points;

(35) $$\tilde{y}_l = F_l\left(\left\{a_1, e_1, i_1, \Omega_1, \omega_1\right\}, \lambda_{\odot}\right),\; l = \overline{1,4}$$

Each of the 4 found solutions ought to be checked by the condition $\beta \le \frac{\pi}{2}$, and 2 of 4 intersection points shall be selected.

Analysing the obtained coefficients for a 4th-order equation, one can see that there are several cases when equation (33) can be significantly simplified in such a way that the solution for the true anomaly can be obtained explicitly.

**Simplification *1.*** $\sin 2\varpi_{\odot} \to 0 \Rightarrow \varpi_{\odot} \to m\frac{\pi}{2},\ m = 0,1,2...,\ \cos 2\varpi_{\odot} \to 1, B, D \to 0$

This solution corresponds to the case when the line connecting the Sun and the planet is located in the vicinity of the apsidal line, as well as in the vicinity of the ascending and descending nodes of the orbit. In this case, eq. (33) is simplified and might be written as

(36) $$\tilde{A}\,\tilde{y}^4 + \tilde{C}\,\tilde{y}^2 + \tilde{E} = 0 .$$

In this case, it is possible to write the following equation depending on the true anomaly:

(37) $$\tilde{y} = \tan\frac{f_{1,\kappa}^{\ in/sc}}{2} = \pm_{\kappa}\sqrt{\frac{\chi e_2^2 + \chi + e_1^2 - 1 \pm_{\kappa} 2\sqrt{\chi\left[e_1^2 + e_2^2(\chi - 1)\right]}}{\chi e_2^2 - \chi + e_1^2 - 2e_1 + 1}},\ \ \kappa = 1,2 .$$

The equation can be rewritten considering Keplerian elements.

(38) $$\tan^2\frac{f_{1,\kappa}^{\ in/sc}}{2} = 1 + \Psi_{\kappa} + \Xi_{\kappa}, \kappa = 1,2\ ,$$

(39) $$\Psi_{\kappa} = 2\frac{\left(1/p_1^2\right)e_1 - 1/p_1^2 + 1/R^2}{\left(1/p^2{}_1\right)e_1^2 - \left(2/p_1^2\right)e_1 + 1/p^2{}_1 - \left(1/R^2\right)\sin^2(i_1)\sin^2(\lambda_{\odot} - \Omega_1)},$$

(40) $$\Xi_{\kappa} = -\frac{2}{R}\frac{\sqrt{\left(1/p^2{}_1\right)e_1^2 + \left(1/p^2{}_1 - 1/R^2\right)\sin^2(i_1)\sin^2(\lambda_{\odot} - \Omega_1) + 1/R^2 - 1/p_1^2}}{\left(1/p^2{}_1\right)e_1^2 - \left(2/p_1^2\right)e_1 + 1/p^2{}_1 - \left(1/R^2\right)\sin^2(i_1)\sin^2(\lambda_{\odot} - \Omega_1)} .$$

It can be shown that $(\Psi_{\kappa} + \Xi_{\kappa}) \sim 10^{-3}$; in this case, the Eq. (39) is simplified, and in the final form, the tangent of half the true anomaly can be expanded in a series in terms of a small parameter $\upsilon = \Psi_{\kappa} + \Xi_{\kappa}$.

(41) $$\tan\frac{f_{1,\kappa}^{\ in/sc}}{2} = \pm_{\kappa}\sum_{j=0}^{\infty}\frac{(-1)^j(2j)!}{(1-2j)(j)!^2(4^j)}\upsilon^j = 1 + \frac{1}{2}\upsilon - \frac{1}{8}\upsilon^2 ... ,\ ,\ \ \kappa = 1,2 .$$

The final solution for the true anomaly is:

(42) $$f_{1,\kappa}^{\ in/sc} = 2\sum_{k=0}^{\infty}\frac{(-1)^k}{2k+1}\tilde{\upsilon}^{2k+1},\ \tilde{\upsilon} = \pm_{\kappa}\sum_{j=0}^{\infty}\frac{(-1)^j(2j)!}{(1-2j)(j)!^2(4^j)}\upsilon^j,\ \ \kappa = 1,2 .$$

Note that in the following case, the two roots were obtained apart from two trivial solutions due to the symmetry of the final solution.

**Simplification *2.*** It can be shown that the quartic equation eq. (33) reduces to a palindromic equation for circular orbits (i.e., $e_1 = 0$). To further avoid uncertainty in $f_1$ for circular orbits, $f_1^{\ in/sc}$ is replaced by $u_1^{\ in/sc}$, where $u_1$ is the angle between the positions of the ascending node and the satellite.

(43) $$\tilde{A}\left(\tilde{z}^2+\frac{1}{\tilde{z}^2}\right)+\tilde{B}\left(\tilde{z}-\frac{1}{\tilde{z}}\right)+\tilde{C}=0,$$

where $\tilde{z}=\tan\frac{u_1^{in/sc}}{2}$.

Using methods described, for example, in [31], the final solution to the above eq. (43) can be written as:

(44) $$\tilde{z}=\tan\frac{u_{1,\kappa}^{in/sc}}{2}=\left(\nu_1 \pm_l \nu_2 \pm_\kappa \nu_3\right),\ \ \kappa=1,2;\ l=1,2,$$

where coefficients in the solution are functions of orbital elements:

$$\nu_1=\frac{e_2^2\chi\sin(2\varpi_\odot)}{2(e_2^2\chi\cos(\varpi_\odot)^2-\chi+1)},\ \nu_2=\frac{2\sqrt{e_2^2\chi^2-e_2^2\chi-\chi^2+2\chi-1}}{2(e_2^2\chi\cos(\varpi_\odot)^2-\chi+1)},\ \nu_3=\frac{\Lambda_2}{2(e_2^2\chi\cos(\varpi_\odot)^2-\chi+1)},$$

$$\Lambda_2=\sqrt{2}\sqrt{e_2^2\chi\left(e_2^2\chi-2\chi+2\right)\cos(2\varpi_\odot)+\sin(2\varpi_\odot)+e_2^4\chi^2-2e_2^2\chi\left(\chi-1\right)}^{1/2}\Lambda_1,$$

$$\Lambda_1=\sqrt{e_2^2\chi-\left(\chi-1\right)}.$$

**Simplification *3*.** $p_1=\sqrt{2}a_2\sqrt{\frac{e_2^2-1}{\cos 2\varpi_\odot-1}}\left(e_1+1\right)\rightarrow\tilde{E}=0$.

In this case, equation (33) takes the following form:

(45) $$\tilde{y}\left(\tilde{A}\tilde{y}^3+\tilde{B}\tilde{y}^2+\tilde{C}\tilde{y}+\tilde{D}\right)=0.$$

The solution to the above equation has real roots. However, due to the complexity of the resulting solution, it is not presented in this paper. Notice that the discussed simplification is valid for orbits with a semilatus rectum and eccentricity in relation to shadow ellipse parameters.

## 1.4. Algorithm for considering changing shadow borders due to the planet's motion

The motion of the planet follows the changing position of the Sun in the heliocentric frame. This causes changes in the shadow boundaries determined by the developed model. The primer estimation of the semi-major axis of the satellite orbit can be made $a_1^*\geq\sqrt[3]{\frac{P_{pl}^2\mu_{pl}}{16\pi^4}\Delta\theta^2}$, where $\Delta\theta$ is the relatively small angular motion of the planet that can be taken between 2 to 25 deg (smaller values for inner planets, higher for outer ones). This value constrains the region in $a_1$ in which method can be used without any corrections. For Earth this value is $a_1^*=67{,}703.89$ km considering $\Delta\theta=2\,\text{deg}$. Hence, for cases when $a_1$ greater than restriction the corrections to the sun's position is required.

To consider the effect of the planet's heliocentric position changing on the final estimation of satellite eclipses duration, one can use the following numerical algorithm:

**Input**: $\tilde{\varepsilon} = 10^{-3}$ is the given accuracy, $\{a_1, e_1, i_1, \Omega_1, \omega_1\}$, $f_1$ at given epoch $t_0$.

**Output**: ${}^{p}f_{+/-}$.

**Perform the following iterations** while given accuracy is reached:

1. Calculate the parameters of the Sun relative to the satellite orbital plane for a given epoch:
   1.1 Sun-orbit angle $\beta_{\odot}$.
   1.2 Sun's position in the orbital plane relative to the periapsis $\varpi_{\odot}$.
2. Determine the parameters of the shadow ellipse $a_2, b_2, e_2$ in the orbital plane for given orbital parameters $\{a_1, e_1, i_1, \Omega_1, \omega_1\}$ and the Sun's orientation ($\beta_{\odot}, \varpi_{\odot}$).

2. Determine the true anomaly at the calculated shadow boundary.

(46)
$$f_{1,+/-} = \tilde{F}_{+/-}\left(\{a_1, e_1, i_1, \Omega_1, \omega_1\}, \lambda_{\odot}\right)$$

3. Calculate the time of flight from the current satellite position $f_1^{\,i}$ to one that corresponds to an entry to shadow $f_{1,+}$.
4. Determine the difference in true anomaly between the current value of the shadow entry point and the previously obtained value.
   The difference in true anomaly can be calculated using the expression:

(47)
$$\Delta f_{1,+} = f_{1,+}{}^{new} - f_{1,+}{}^{old},$$

where $f_{1,+}{}^{new}$ is the true anomaly of the satellite at the new shadow boundary and $f_{1,+}{}^{old}$ is the true anomaly of the satellite at the previously calculated shadow boundary (on the first iteration, it can be assumed to be zero).

5. Check the condition for stopping $\Delta f_{1,+} < \tilde{\varepsilon}$; if not, then continue with updated values for the satellite current position and the sun's position.
   5.1 Recalculate satellite position:

(48)
$$f_1^{\,i+1} = \begin{cases} f_1^{\,i} + \Delta f_{1,+}, i_1 \le 90\,\text{deg} \\ f_1^{\,i} - \Delta f_{1,+}, i_1 > 90\,\text{deg} \end{cases}.$$

5.2 Recalculate the planet position assuming that the planet is in an elliptical orbit around the Sun:

(49)
$$M^{S,i+1} = M^{S,i}{}_0 + n_S \frac{t_i - t_0}{T_S},$$

where $n_S$ is the mean motion of the planet in its heliocentric orbit, $T_S$ is the planet's heliocentric period, $t_i, M^{S,i+1}$ are the current time and anomaly on the satellite eclipses, and $t_0, M^{S,i}{}_0$ are the time and mean anomaly of the Sun at the beginning of the calculation. Note that that step can be changed by using precise ephemeris, for example, DE441 by JPL[4], which allows one to obtain the planet's position with much more accuracy than the Keplerian approximation.

7. Repeat steps until the condition in step 5 is true.

# 2. Results

## 2.1 Tests for accuracy in the determination of eclipse duration for different sets of orbits.

In this section, the comparison of the analytical model established in this research with the numerical model [3,32-33] is provided. The estimation of the shadow time obtained in the proposed approach might be compared with the numerical estimation[5]:

(50)
$$\varepsilon = \frac{|t_{Analytic} - t_{Numeric}|}{t_{Numeric}} 100\%,$$

where $t_{Analytic}$, $t_{Numeric}$ are the time of the satellite's stay in the shadow of a spherical planet, calculated analytically and numerically.

All tests in this paragraph were performed for Earth satellites whose test orbits are presented in Table 1. For all initial conditions in the table, semimajor axes $a_1$ are taken from 10,000 to 270,000 km.

*Table 1. Set of orbits for testing the established eclipse model*

| *N* | *Date, UTC* | $e_1$ | $i_1$, deg | $\Omega_1$, deg | $\omega_1$, deg | $f_1$, deg | $\mu_{pl}$, $km^3/s^2$ |
|---|---|---|---|---|---|---|---|
| 1 | Sep 5, 2032 00:00:00.000 | 0.1 | *0*<br>*30*<br>*60*<br>*90* | 0 | 0 | 0 | 398600.4415 |
| 2 | | 0.35 | *0*<br>*30*<br>*60*<br>*90* | | | | |
| 3 | | 0.6 | *0*<br>*30*<br>*60*<br>*90* | | | | |

[4] https://ssd.jpl.nasa.gov/planets/eph_export.html (Accessed Sep 9, 2023).

[5] The algorithm used is the one from Montenbruck [2, pp. 80-81].

| | | |
|---|---|---|
| 4 | 0.85 | *0*<br>*30*<br>*60*<br>*90* |

Assuming that the motion of the satellite begins with $f_1 = 0\ \text{deg}$, the eclipse duration is calculated for one revolution of the satellite around the planet, i.e., $f_1 \in [0;360)$.

In Fig. 7, the results for calculating the eclipse duration using SEEM are provided for all of the abovementioned satellite orbit sets.

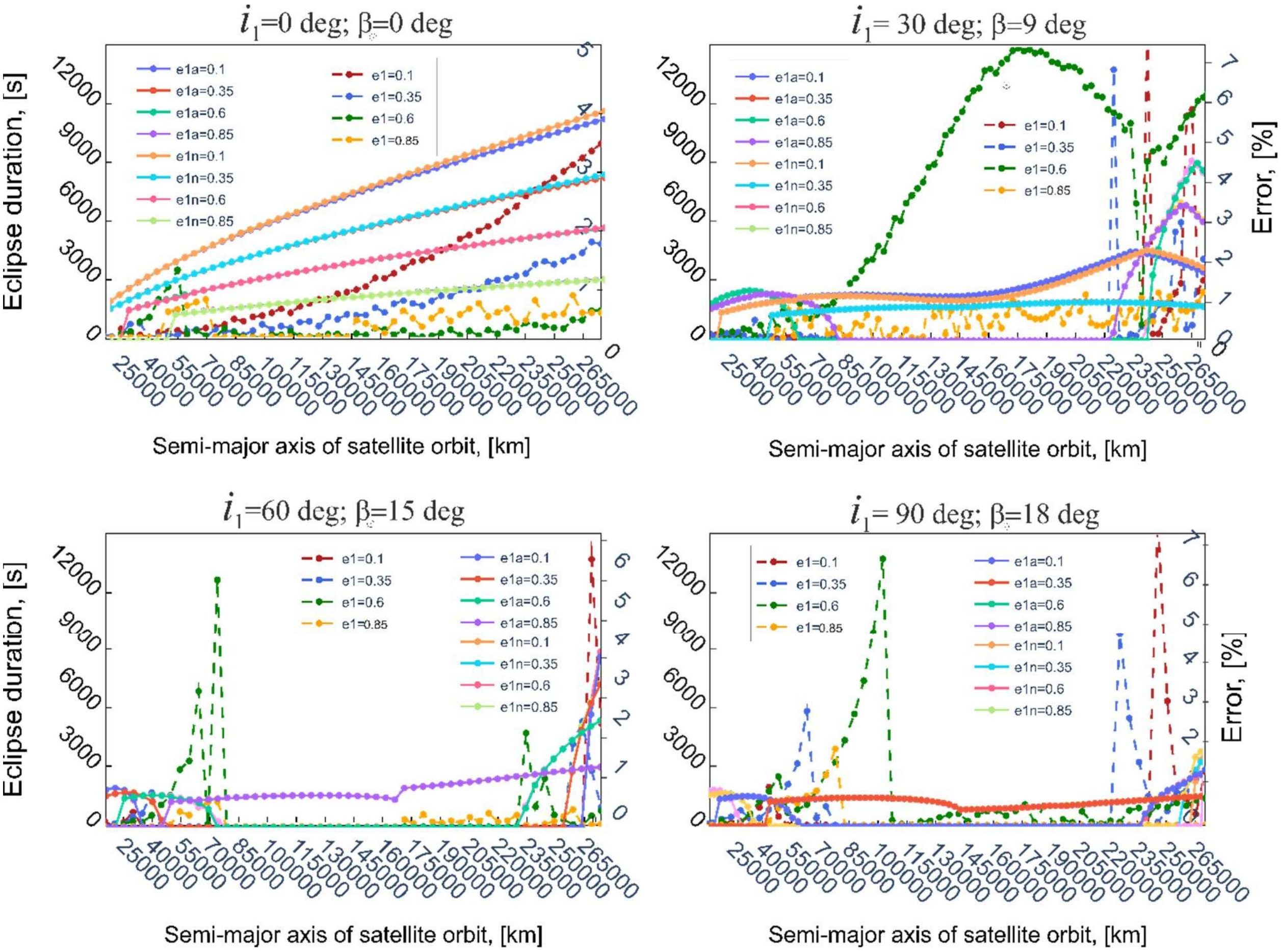

Figure. 7 Eclipse duration vs. semimajor axis of studied orbits for Sep 5, 2032 00:00: UTC. Designations in plots “e1a” and “e1n” differentiate the curves obtained analytically and numerically, respectively. Dashed lines indicate error estimation while solid lines correspond to actual occultation duration.

Analyzing the maximum error (Fig. 7), it should be noted that the lowest accuracy in determining the shadowing duration is in the borderline cases, when there is a transition to the critical value of $a_1$, after which the spacecraft orbit is always out of the planet's for chosen $i_1$.

It turns out to be interesting to plot the dependence of the relative error of determining the duration of the spacecraft in the penumbra region, for example, for the near-plane case ($i_1$=0 deg). In Fig. 8, the mentioned error is drawn with respect to the semimajor axis of the orbit set in the study. Additionally, for this calculation, the same set of the satellite orbits with Keplerian elements defined in Table 2 is used.

*Table 2. Set of orbits for testing the established eclipse model*

| *Date, UTC* | $e_1$ | $i_1$, deg | $\lambda_1$, deg | $f_1$, deg | $\mu_{pl}$, $km^3/s^2$ |
|---|---|---|---|---|---|
| Sep 5, 2032 00:00:00.000 | 0<br>*0.25*<br>*0.75*<br>*0.9* | *0* | *0* | *0* | 398600.4415 |

Both considered cases with sets for Fig. 7 and Fig. 8 show that the SEEM demonstrates satisfactory results in terms of major coverage in the eccentricity and semimajor axis range.

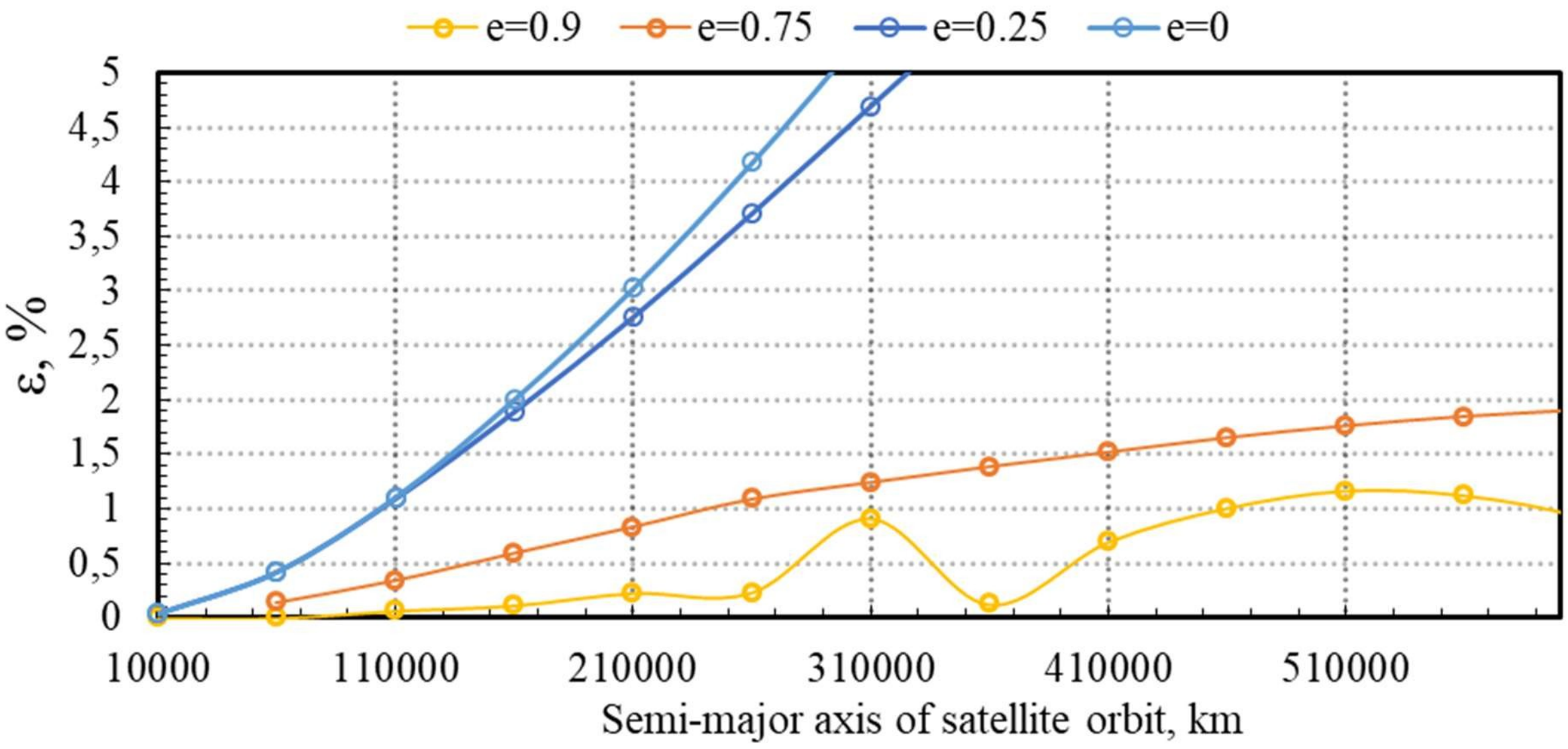

*Figure. 8 Error in determining the eclipses duration vs. semimajor axis of studied orbits for plane case scenario (1st set) for the date Sep 5, 2032 00:00UTC.*

Fig. 8 shows that accurate calculations of the satellite eclipses can be achieved for semimajor axes less than 600,000 km, which are nearly identical to the numerical results for large eccentricities (with an error of less than 2%). This is because as eccentricity decreases, the direction towards the periapsis becomes less distinct, resulting in errors as the semimajor axis of the orbit increases. It is important to note that for semimajor axes below 100,000 km, the aforementioned effects are negligible, and SEEM demonstrates good results in determining shadowing for this region across all ranges of eccentricity studied.

## 2.2. Tests for hyperbolic orbits

This case is particularly interesting for the satellite flyby of a planet, during landing on its surface or in the process of a gravity-assist manoeuvre. The developed method allows convenient determination of the time a satellite spends in the shadow of a planet during a gravity assist manoeuvre when the satellite's orbit within the sphere of influence is determined by the asymptotic approach and departure velocities from the planet obtained by an analytical method.

As an example, let us consider the passage of a satellite through the Earth's sphere of influence along a hyperbolic trajectory in the ecliptic plane (Table 3, $N$=1) and with inclination of 40 degrees to it (Table 3, $N$=2). The error magnitude and the eclipse duration value were estimated for changing the longitude of the perigee within the range of 0 to 360 degrees for the planar case and within the range of 0 to 360 degrees of longitude of the ascending node for the spatial case.

Table 3. Parameters of the test orbits

| $N$ | *Date, UTC* | $a_1$, km | $e_1$ | $i_1$, deg | $\lambda_1^*$, deg | $f_1$, deg | $\mu_{pl}$, $km^3/s^2$ |
|---|---|---|---|---|---|---|---|
| 1 | Sep 5, 2032 | -25000 | 1.5 | *0* | 0 | 0 | 398600.4415 |
| 2 | | | | *45* | 0* | 0 | |

Notice: *in this case $\omega_1$ is set

Let us show the dependence of the duration of the eclipse as well as an error in determining it using the SEEM compared to the numerical model of the satellite on the initial longitude of the perigee of the hyperbolic orbit in Fig. 9.

In this case, the error of the eclipse duration determination may reach 12%. The main reason for the growth of the eclipse duration error is that the satellite orbit is turned relative to the planet-Sun line so that the shadow boundary falls on the part of the orbit that is close to a straight line (i.e., where the true anomaly of the satellite is $f_1 \to f_1^* = \arccos(-1/e_1)$), so that a small error in determining the shadow boundary leads to an increasing error in determining the eclipse duration. Fig. 9*a, b* shows that for the considered example. Consequently, the method based on an analytical approach can be used to determine the time a satellite spends in the shadow of a planet during a gravity-assist manoeuvre, but other factors, such as the orientation of the orbit, should also be taken into account.

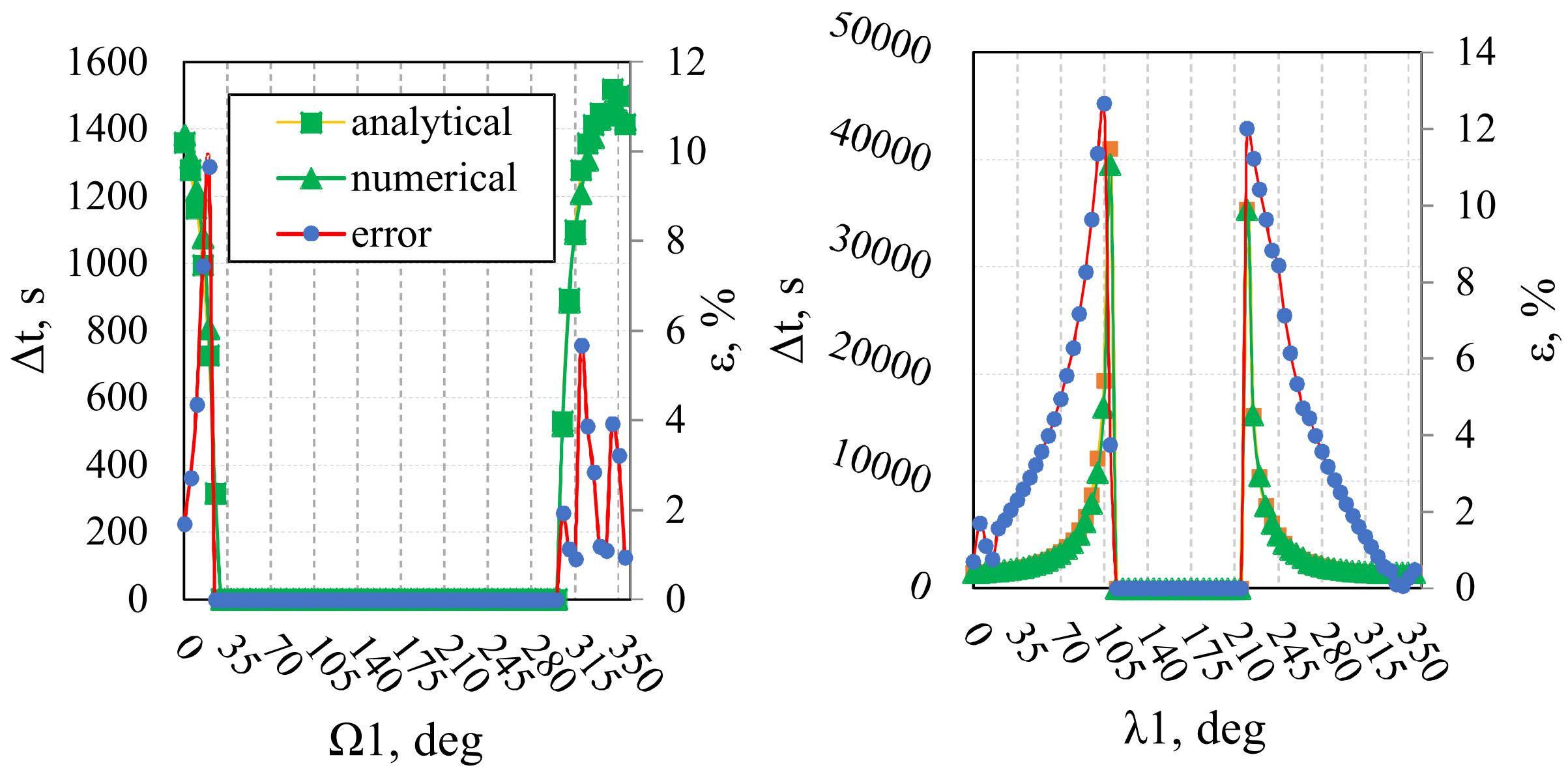

(a) (b)

Figure. 9 Eclipse duration and estimated error vs. (a) longitude of periapsis for $i_1$=0 deg; (b) longitude of ascending node for $i_1$ = 45 deg.

### 2.3 Tests for non-Earth satellites

It is highly interesting to explore the application of the SEEM to calculate occultation events for non-Earth satellites. As an example, consider the satellite "Mars Orbiter Mission", for which the data on its occultation events are publicly available in an article authored by Srivastava et al. [20]. To perform the calculations, one can utilize the satellite's positions and velocities as provided in the aforementioned article and adjust them for the technique in this paper by converting them from the original EME2000 reference frame (as mentioned in [20]) to the Martian frame, which aligns with the Mars heliocentric plane. These satellite positions are presented in Table 4.

In Table 5, the results for calculating the time of occultation events for Martian satellites specifically entry into and exit from the penumbra region are shown. The results were compared with numerical calculations in the software General Mission Analysis Tool (GMAT) and System Tool Kit (STK) (from ref.[20]) using high-precision propagation models and actual data provided in an article written by Srivastava et al. [20]. The designations of "no iter" and "iter included" in the first and second columns of Table 5 refer to sec. 1.4, in which the algorithm for including the Sun's position changing to determine the shadow borders is presented. Basically, "no iter" means that no iterations described in section 1.4 are applied and vica verse for "iter included".

Table 4. Initial state vectors and orbital elements for the Mars Orbiter Mission from ref. [20]. Initial state vectors are given in EME2000 coordinates and orbital elements corresponding to the Mars-Sun plane.

| | Date * | Time | State vector (in EME2000 frame) | | | | | |
|---|---|---|---|---|---|---|---|---|
| 1 | 10 Oct 2014 | 20:15:00.000 | $X_{EME}$, km | $Y_{EME}$, km | $Z_{EME}$, km | $Vx_{EME}$, km/s | $Vy_{EME}$, km/s | $Vz_{EME}$, km/s |
| | | | 28811.51 | 48031.76 | 35377.10 | 0.0816 | -0.3610 | -0.2512 |
| | | | Keplerian elements (corresponding to Mars-Sun plane) | | | | | |
| | | | $a_1$, km | $e_1$ | $i_1$, deg | $\Omega_1$, deg | $\omega_1$, deg | $f_1$, deg |
| | | | 39189.63 | 0.903 | 167.96 | 168.86 | 275.49 | 189.41 |
| | Date | Time | State vector (in EME2000 frame) | | | | | |
| 2 | 18 Oct 2014 | 20:35:00.000 | $X_{EME}$, km | $Y_{EME}$, km | $Z_{EME}$, km | $Vx_{EME}$, km/s | $Vy_{EME}$, km/s | $Vz_{EME}$, km/s |
| | | | 27702.40 | 52199.72 | 38643.80 | 0.1326 | -0.2637 | -0.1822 |
| | | | Keplerian elements (corresponding to Mars-Sun plane) | | | | | |
| | | | $a_1$, km | $e_1$ | $i_1$, deg | $\Omega_1$, deg | $\omega_1$, deg | $f_1$, deg |
| | | | 39192.87 | 0.904 | 167.67 | 169.55 | 276.61 | 186.29 |

**Dates given in Coordinated Universal Time (UTC) format*

Table 5. Comparison of different models for calculating the parameters of the satellite of Mars on the border of shadow

| | Shadow ellipse model SEM (no iter) | Shadow ellipse model SEM (iter included) | GMAT | STK (in ref. [20]) | Srivastava (actual from measurements) [20] |
|---|---|---|---|---|---|
| 11 Oct 2014 | | | | | |
| Penumbra entry time (UTC) | 15:09:46 | 15:09:34 | 15:09:34 | 15:09:33 | 15:09:24 |
| True anomaly of satellite at penumbra entry, deg | 282.37 | 282.05 | 282.07 | 282.07 | 281.82 |
| Penumbra exit time (UTC) | 15:39:51 | 15:39:46 | 15:39:42 | 15:39:41 | 15:39:55 |
| True anomaly of satellite at penumbra exit, deg | 17.51 | 17.13 | 16.94 | 16.91 | 17.85 |
| Duration of satellite staying in penumbra, s | 1805.55 | 1811.82 | 1808.45 | 1808 | 1831 |
| $\varepsilon$, % | 1.38 | 1.04 | 1.23 | 1.25 | - |
| 19 Oct 2014 | | | | | |
| Penumbra entry time (UTC) | 19:28:21 | 19:28:05 | 19:28:05 | 19:28:03 | 19:27:55 |
| True anomaly of satellite at penumbra entry, deg | 278.66 | 278.26 | 278.27 | 278.28 | 278.08 |
| Penumbra exit time (UTC) | 19:59:36 | 19:59:29 | 19:59:26 | 19:59:24 | 19:59:46 |
| True anomaly of satellite at penumbra exit, deg | 13.73 | 13.26 | 13.08 | 13.05 | 14.57 |
| Duration of satellite staying in penumbra, s | 1875.62 | 1884.61 | 1881.21 | 1881 | 1911 |
| $\varepsilon$, % | 1.85 | 1.38 | 1.56 | 1.57 | - |

The results presented in Table 5 show that despite being developed for rapid and preliminary calculations of occultation events, SEEM provides enough accuracy for estimation compared with numerical estimators and real data from satellites. The average error for the two considered dates with events on Oct 11 and 19, 2014, does not exceed even 2% compared to actual data with no corrections to the nonsphericity of Mars and even without an iteration procedure (sec. 1.4) for adapting the Sun's position.

## 2.4 Tests for low Earth orbit satellites

This section presents a comprehensive evaluation of the developed SEEM model applied to eclipse prediction for low-Earth orbit (LEO) satellites. The main objective of these tests was to assess the accuracy and reliability of the SEEM model by comparing its results with actual data obtained from satellites. To carry out these tests GMAT software is utilized providing a robust platform for conducting various simulations and analyses. However, in the GMAT tests, unlike in the previous section, only two body problems are solved, which are indicated in Table 7. Additionally, data presented in reference [16] for comparison purposes. This section attempts to demonstrate the effectiveness and applicability of the SEEM model in accurately predicting and modelling the environmental conditions experienced by Earth satellites. Initial states for LEO satellites that are taken from [16] are provided in Table 6 in the International Celestial Reference Frame (ICRF), and Keplerian elements are given in the Ecliptic frame.

Table 7 presents the results of the comparison of the utilized models with the actual satellite data for LEO IRS satellites OCN-2 and CAR-2A. The models evaluated in the table include SEEM, STK (high-precision model), SEEM TBP, SEEM (corr), and GMAT TBP (two-body problem). This comparative analysis provides valuable insights into the accuracy and effectiveness of the models in accurately representing the observed data. By examining the values in Table 7, readers can gain a comprehensive understanding of the performance and reliability of the models in relation to the actual data. This comparison serves as a crucial step in evaluating the effectiveness and applicability of the models in accurately capturing the real-world environmental conditions for Earth satellites.

Table 6. Initial state vectors and orbital elements for the Indian Research Satellites (IRS) from *ref. [16]*

| Satellite | Date | State vector (in ICRF frame, Earth centered) | | | | | |
|---|---|---|---|---|---|---|---|
| OCN-2 | Nov 22, 2013 (at 0UTC) | $X_{ICRF}$, km | $Y_{ICRF}$, km | $Z_{ICRF}$, km | $Vx_{ICRF}$, km/s | $Vy_{ICRF}$, km/s | $Vz_{ICRF}$, km/s |
| | | 3728.863 | 5741.984 | 1890.266 | −0.14028 | −2.27027 | 7.13946 |
| | | Keplerian elements (corresponding to ecliptic plane) | | | | | |
| | | $a_1$, km | $e_1$ | $i_1$, deg | $\Omega_1$, deg | $\omega_1$, deg | $f_1$, deg |
| | | 7105.95 | 0.00127 | 86.049 | 58.533 | 64.438 | 291.111 |
| | Date | State vector (in ICRF frame, Earth centered) | | | | | |
| CAR-2A | Nov 26, 2013 (at 0 UTC) | $X_{ICRF}$, km | $Y_{ICRF}$, km | $Z_{ICRF}$, km | $Vx_{ICRF}$, km/s | $Vy_{ICRF}$, km/s | $Vz_{ICRF}$, km/s |
| | | −1236.77 | −1683.742 | 6685.318 | −6.59988 | −3.05537 | −1.9969 |
| | | Keplerian elements (corresponding to ecliptic plane) | | | | | |
| | | $a_1$, km | $e_1$ | $i_1$, deg | $\Omega_1$, deg | $\omega_1$, deg | $f_1$, deg |
| | | 7000.71 | 0.00095 | 77.087 | 27.516 | 215.918 | 238.848 |

Note that the distinction between the SEEM TBP and SEEM (corr) is that the latter is applied to corrected orbital elements for the calculated initial orbit of OCN-2. The intermediate Keplerian elements for the orbit were obtained after propagating the initial state vector from Table in the high-precision model to Nov 22, 2013 04:02:11 (UTC) using GMAT and are $a_1 = 7107.081\,\text{km}$ $e_1 = 0.001232$ $i_1 = 86.10361\,\text{deg}$, $\Omega_1 = 58.69007\,\text{deg}$, and $\omega_1 = 95.8662\,\text{deg}$. The corrected results for satellite CAR-2A were obtained for orbit $a_1 = 7007.561\,\text{km}$ $e_1 = 0.000667$ $i_1 = 77.16034\,\text{deg}$, $\Omega_1 = 27.9251\,\text{deg}$, $\omega_1 = 120.3852\,\text{deg}$ (elements on Nov 26, 2013 at 09:51:39).

Table 7. Comparison of SEEM application for LEO IRS satellites OCN-2 and CAR-2A from [16].

| | *N* of orbtis | Model used | Penumbra entry, UTC | Umbra entry, UTC | Umbra exit, UTC | Penumbra exit, UTC |
|---|---|---|---|---|---|---|
| OCN-2 | 4 | STK (high precision model) [16] | 4:41:36 | 4:41:45 | 5:16:40 | 5:16:49 |
| | | Actual data [16] | 4:41:39 | 4:41:51 | 5:16:36 | 5:16:52 |
| | | SEEM TBP | 4:41:41 | 4:41:50 | 5:16:50 | 5:16:59 |
| | | SEEM (corr) | 4:41:36 | 4:41:45 | 5:16:46 | 5:16:55 |
| | | GMAT TBP | 4:41:44 | 4:41:53 | 5:16:46 | 5:16:56 |
| | 5 | STK (high precision model) [16] | 6:20:55 | 6:21:04 | 6:55:59 | 6:56:08 |
| | | Actual data [16] | 6:20:54 | 6:21:07 | 6:55:56 | 6:56:08 |
| | | SEEM TBP | 6:21:03 | 6:21:12 | 6:56:12 | 6:56:21 |
| | | SEEM (corr) | 6:20:59 | 6:21:08 | 6:56:09 | 6:56:18 |
| | | GMAT (two-body) | 6:21:06 | 6:21:14 | 6:56:07 | 6:56:17 |
| | 6 | STK (high precision model) [16] | 8:00:14 | 8:00:23 | 8:35:18 | 8:35:27 |
| | | Actual data [16] | 8:00:18 | 8:00:26 | 8:35:11 | 8:35:32 |
| | | SEEM TBP | 8:00:24 | 8:00:33 | 8:35:33 | 8:35:42 |
| | | SEEM (corr) | 8:00:21 | 8:00:30 | 8:35:31 | 8:35:40 |
| | | GMAT TBP | 8:00:27 | 8:00:36 | 8:35:29 | 8:35:38 |
| CAR-2A | 7 | STK (high precision model) [16] | 10:42:59 | 10:43:11 | 11:14:55 | 11:15:06 |
| | | Actual data [16] | 10:42:59 | 10:43:11 | 11:14:52 | 11:15:08 |
| | | SEEM TBP | 10:40:54 | 10:41:06 | 11:12:57 | 11:13:09 |
| | | SEEM (corr) | 10:42:46 | 10:42:58 | 11:14:51 | 11:15:03 |
| | | GMAT TBP | 10:40:57 | 10:41:08 | 11:12:52 | 11:13:04 |
| | 8 | STK (high precision model) [16] | 12:20:26 | 12:20:38 | 12:52:22 | 12:52:33 |
| | | Actual data [16] | 12:20:12 | 12:20:36 | 12:52:25 | 12:52:37 |
| | | SEEM TBP | 12:18:04 | 12:18:16 | 12:50:06 | 12:50:18 |
| | | SEEM (corr) | 12:20:05 | 12:20:17 | 12:52:09 | 12:52:21 |
| | | GMAT TBP | 12:18:07 | 12:18:18 | 12:50:02 | 12:50:13 |

| | | | | | |
|---|---|---|---|---|---|
| | STK (high precision model) [16] | 13:57:53 | 13:58:05 | 14:29:49 | 14:30:00 |
| 9 | Actual data [16] | 13:57:45 | 13:58:05 | 14:29:54 | 14:30:06 |
| | SEEM TBP | 13:55:15 | 13:55:26 | 14:27:16 | 14:27:28 |
| | SEEM (corr) | 13:57:24 | 13:57:35 | 14:29:27 | 14:29:39 |
| | GMAT TBP | 13:55:17 | 13:55:28 | 14:27:11 | 14:27:22 |

** TBP = two-body problem, results obtained for Keplerian orbits from Table 3; (corr) designates results obtained used corrected orbital elements for osculation.*

Table 7 shows that the difference between the modelled times of the umbra and penumbra boundary crossing and the real data increases and amounts to minutes in case the values are obtained in the SEEM TBP model (i.e., including correction for the planet motion but using the information on the Keplerian orbit of the satellite) as well as in the GMAT TBP model, vice versa for cases of applying the SEEM (corr) and STK models when the difference between the real data and the modelled one in maximum of 10*s*.

However, even in the absence of correction of initial conditions in the SEEM model, the time of passing the shadow boundaries over several revolutions (approximately 5 revolutions in the examples in Table 7) is determined with an accuracy of up to seconds in comparison with real data and modelled in high-precision models of satellite motion. Thus, the application of model SEEM TBP is restricted to no more than a few revolutions for LEO satellites and tightened for satellites orbiting on high elliptical orbits with much greater period than discussed.

## CONCLUSION

This paper presents a simplified technique for rapid engineering calculation of the boundaries of the shadow region of a planet's orbit. The equations used to determine the shadow boundaries are given for the cases when the satellite moves in a Keplerian orbit in the same plane as the Sun and when their motion takes place in different planes. The algorithm accounts for the planet's motion during the satellite orbit. A comparison is made between the calculated time of the satellite's stay in the shadow of the planet using the proposed approach and the numerical integration of the equations of motion.

1. It is shown that the main advantage of the proposed approach to an analytical solution to find the occultation events is that it is shown that the shadow of the planet can be projected in the orbital plane and described as an ellipse with its semimajor and semiminor axes. The representation of these as a function of the orbital elements and the Sun's orientation parameters $a_2 = \tilde{f}(K)$, $b_2 = \tilde{f}(K)$, $K = \{a_1, e_1, \lambda_\odot, i_1, \Omega_1, \omega_1\}$ gives the possibility to construct any analytical or semi-analytical approaches, if the considered problem can be reduced to the geometric one.

2. The analytical model presented in the article allows one to obtain a proper estimate for the duration of eclipses for satellite elliptical orbits up to $a_1 \leq 150{,}0000$ km (for Earth orbiting satellites). The obtained model can also be applied with some accuracy for $a_1 > 150{,}000$ km; however, one should note that there is no practical sense of such application for Keplerian orbits because in the case of, for example, near-Earth satellites, the perturbations caused by gravitational effects of the Sun and Moon lead to a significant deviation of the initial orbit from the initial Keplerian orbit.
3. The given model can also be applied to calculate hyperbolic orbits, but the error in determining the duration of the eclipse can reach 10-12%, which is caused because the boundaries of the shadow appear on the section of the motion of the spacecraft close to a straight line, so that any small error in determining the boundaries of the beginning and end of the eclipse leads to a large error in determining the duration of the eclipse.
4. The results shown in the case of applying the developed SEEM model for calculating eclipses for the Martian satellite and comparing it to actual data displayed good accuracy regardless of whether the iteration on the Sun position was applied (1.04%/1.38% for the 1st case and 1.38%/1.85% for the 2nd case). It is particularly interesting to note that the best fit of the SEEM model determines the boundary of the spacecraft's exit from the planet's shadow.
5. The comparison between the SEEM model and the actual satellite data, shows that predicted times of entry/exit to and out of penumbra/umbra regions are determined with an accuracy of tens of seconds in case if osculation of LEO orbit satellites are taken into account on every revolution of satellite around the Earth and 1-2 min. if correction is not applied during significant time intervals (about one or two hours for LEO satellites).

## DECLARATION OF COMPETING INTEREST.

The authors have no competing interests to declare that are relevant to the content of this article.

## AUTHOR CONTRIBUTIONS.

All authors contributed to the study conception and design. Material preparation, data collection and analysis were performed by Vladislav Zubko and Andrey Belayev. The first draft of the manuscript was written by Vladislav Zubko, and all authors commented on previous versions of the manuscript. All the authors have read and approved the final manuscript.

## BIBLIOGRAPHY

[1] B. Neta, D. Vallado, On satellite umbra/penumbra entry and exit positions, J. Astronaut. Sci. 46 (1998) 91–103.

[2] J.R. Wertz, Spacecraft attitude determination and control, Springer Science \& Business Media, 2012.

[3] O. Montenbruck, E. Gill, Satellite Orbits: Models, Methods, Applications, Springer, Berlin, Germany, 2005.

[4] L.G. Stoddard, Eclipse of an Artificial Earth Satellite, Astronaut. Sci. Rev. 3 (1961) 9–16.

[5] G.B. Patterson, Graphical method for prediction of time in sunlight for a circular orbit, ARS J. 31 (1961) 441–442.

[6] G.W. Peckham, The orbital shadow time of an earth satellite, Air Force Institute of Technology, 1960.

[7] F.T. Geyling, H.R. Westerman, Introduction to orbital mechanics, 1971.

[8] P. Escobal, Methods of orbit determination., John Wiley & Sons, 1965.

[9] L.D. Mullins, Calculating satellite umbra/penumbra entry and exit positions and times, J. Astronaut. Sci. 39 (1991) 411–422.

[10] A. V Dobroslavskiy, On Estimating the Average Stay of an Artificial Satellite in the Area of the Earth's Shadow while Moving in the Ecliptic Plane, Cosm. Res. 58 (2020) 501–507.

[11] J. Zhang, K. Wang, B. Yan, L. Wang, Eclipse analysis for small-eccentricity orbits using analytical model, Adv. Sp. Res. 70 (2022) 2323–2333.

[12] S. Adhya, A. Sibthorpe, M. Ziebart, P. Cross, Oblate earth eclipse state algorithm for low-earth-orbiting satellites, J. Spacecr. Rockets. 41 (2004) 157–159.

[13] M. Nugnes, C. Colombo, others, A New Analytical Method for Eclipse Entry/Exit Positions Determination Considering a Conical Shadow and an Oblate Earth Surface, in: 2022 AAS/AIAA Astrodyn. Spec. Conf., 2022: pp. 1–19.

[14] D. Vokrouhlický, P. Farinella, F. Mignard, Solar radiation pressure perturbations for Earth satellites: IV. Effects of the Earth's polar flattening on the shadow structure and the penumbra transitions, Astron. Astrophys. 307 (1996).

[15] V.K. Srivastava, M. Pitchaimani, B.S. Chandrasekhar, others, Eclipse prediction methods for LEO satellites with cylindrical and cone geometries: a

comparative study of ECSM and ESCM to IRS satellites, Astron. Comput. 2 (2013) 11–17.
[16] V.K. Srivastava, A. Ashutosh, M. V. Roopa, B.N. Ramakrishna, M. Pitchaimani, B.S. Chandrasekhar, Spherical and oblate Earth conical shadow models for LEO satellites: Applications and comparisons with real time data and STK to IRS satellites, Aerosp. Sci. Technol. 33 (2014). https://doi.org/10.1016/j.ast.2014.01.010.
[17] Y.J. Song, B.Y. Kim, The effect of the Earth's oblateness on predicting the shadow conditions of a distant spacecraft: Application to a fictitious lunar explorer, Adv. Sp. Res. 57 (2016). https://doi.org/10.1016/j.asr.2015.09.028.
[18] M.N. Ismail, A. Bakry, H.H. Selim, M.H. Shehata, Eclipse intervals for satellites in circular orbit under the effects of Earths oblateness and solar radiation pressure, NRIAG J. Astron. Geophys. 4 (2015) 117–122.
[19] V. Arya, R. Woollands, J.L. Junkins, Indirect based shadow modelling with warm-up time for orbit transfers, in: Astrodyn. Spec. Conf., Univelt, 2022: p. AAS 22-054.
[20] V.K. Srivastava, J. Kumar, S. Kulshrestha, B.S. Kushvah, M.K. Bhaskar, S. Somesh, M. V. Roopa, B.N. Ramakrishna, Eclipse modeling for the Mars Orbiter Mission, Adv. Sp. Res. 56 (2015). https://doi.org/10.1016/j.asr.2015.04.025.
[21] C. Hubaux, A. Lemaître, N. Delsate, T. Carletti, Symplectic integration of space debris motion considering several Earth's shadowing models, Adv. Sp. Res. 49 (2012). https://doi.org/10.1016/j.asr.2012.02.009.
[22] M. Gupta, K. Howell, Cislunar Eclipse Mitigation Strategies for Resonant Periodic Orbits, in: 2023 AAS/AIAA Astrodyn. Spec. Conf., Big Sky, Montana, 2023: p. AAS 23-376.
[23] Y. Zhang, X. Wang, K. Xi, Z. Li, Comparison of shadow models and their impact on precise orbit determination of BeiDou satellites during eclipsing phases, Earth, Planets Sp. 74 (2022) 126.
[24] S.Z. Fixler, Umbra and penumbra eclipse factors for satellite orbits, AIAA J. 2 (1964) 1455–1457.
[25] J.F. Traub, Iterative methods for the solution of equations, American Mathematical Soc., 1982.
[26] M.S. Petković, B. Neta, L.D. Petković, J. Džunić, Multipoint methods for solving nonlinear equations: A survey, Appl. Math. Comput. 226 (2014) 635–660.
[27] M. Petkovic, B. Neta, L. Petkovic, J. Dzunic, Multipoint Methods for Solving Nonlinear Equations, Academic Press, 2013.

[28] R.H. Battin, An Introduction to the Mathematics and Methods of Astrodynamics, Revised Edition, American Institute of Aeronautics and Astronautics, Reston ,VA, 1999. https://doi.org/10.2514/4.861543.
[29] D.A. Vallado, Methods of Astrodynamics, a Computer Approach, 1991.
[30] R.R. Bate, D.D. Mueller, J.E. White, W.W. Saylor, Fundamentals of astrodynamics, 2nd editio, Dover Publications, Garden City, New York, 2020.
[31] Chávez-Pichardo, M.; Martínez-Cruz, M.A.; Trejo-Martínez,A.; Vega-Cruz, A.B.; Arenas-Resendiz, T. On the Practicality of the Analytical Solutions for all Third- and Fourth-Degree Algebraic Equations with Real Coefficients. Mathematics. 11 (2023) 1447. https://doi.org/10.3390/math11061447
[32] O. Montenbruck, E. Gill, F. Lutze, Satellite orbits: models, methods, and applications, Appl. Mech. Rev. 55 (2002) B27--B28.
[33] G.R. Hintz, Fundamentals of Astrodynamics, in: Orbital Mech. Astrodyn., Springer International Publishing, Cham, 2015: pp. 1–21. https://doi.org/10.1007/978-3-319-09444-1_1.

## Appendix A

**Statement 1.** The shape of the shadow in the orbit plane can be described by a second-order curve.

**Proof.** Let the motion of the satellite take place around a celestial body having a spherical shape along a perfect Keplerian trajectory. Let the rays of sunlight propagate exclusively along radial directions. Relativistic effects are neglected. The shadow formed as a result of the obstruction of sunlight rays by a spherical planet can be described by two cones within the framework of the conic model. The shadow cone (umbra) with its apex lying on the planet-Sun line in the direction from the Sun limits the area of full shadow. The cone of the penumbra, whose apex lies on the same line in the direction to the Sun, includes the region of full and partial shadow. Both cones have a common axis of rotation, which coincides with the planet-sun direction and is determined by the half-angle $\alpha_p, \alpha_u$ of their directrix to this axis.

The satellite moves along the Keplerian trajectory, which in the framework of the two-body problem means motion in an invariable plane, which is equivalent to fulfilment of the condition $\|\mathbf{r}(t) \times \mathbf{V}(t)\| = 0$ for all positions and velocities of the satellite taken at the time *t*.

If the above statement is true, then the satellite plane passing through the centre of the celestial body intersects the penumbra or umbra cone at an angle $\beta_\odot$ to the axis of the cone. It is known from the projective geometry that the section of a cone by a plane intersecting its axis at an angle of $\beta_\odot$ generates a second-order curve in this plane, the eccentricity ($e_2$) of which can be determined from the known geometrical relations $e_2 = \cos\beta_\odot / \cos\alpha_p$, since the denominator is $\cos\alpha_p \underset{\alpha_p \to 0}{\to} 1$, by virtue of relations (1), and the angle $\delta$ is determined from the spherical triangle in section 1.1 (2), then $|\beta_\odot| = |-(\lambda_\odot - \Omega_1)i_1|$, then $|(\lambda_\odot - \Omega_1)i_1| > 0 \Leftrightarrow (\lambda_\odot - \Omega_1) \neq 0 \wedge i_1 \neq 0$, whence it follows that the shadow in such a case will be a second-order curve (ellipse) in the plane of the satellite orbit $e_2 < 1$; in the case where $|(\lambda_\odot - \Omega_1)i_1| = 0 \Leftrightarrow (\lambda_\odot - \Omega_1) = 0 \vee i_1 = 0$, the second-order curve degenerates into a pair of intersecting straight lines. The statement has been proved.

## Appendix B. Obtaining closed-form analytical solution for satellite elliptical orbits in near planar cases

Section 1.1 presents a method for calculating the duration of eclipse of a spacecraft by a celestial body in cases when the eclipse is total (umbra) and partial (penumbra). In this method, one of the key features is the representation of the shadow as a second-order curve defined in the orbital plane, for which the semimajor axis and eccentricity are determined geometrically. The determination of these parameters for cases where the Sun is close to the orbital plane, i.e. $\beta_\odot \to 0$, presents additional difficulties because of the need to redefine the coefficients (9) and solve two second-degree equations (11) and then define intersection points between the satellite orbit and the shadow .

Then, the best approach in this case is to introduce a much simpler solution then in sec. 1.1 when $\beta_\odot \to 0$. The foundation for this is the representation of shadow as a pair of intersecting lines defined in a satellite orbital plane (Fig. A.1). In this case, it is easy to solve the problem and obtain the required intersection points of the lines, i.e., sunlight rays, with the satellite orbit.

The assumptions for establishing this solution are the same as those described in section 1.1.

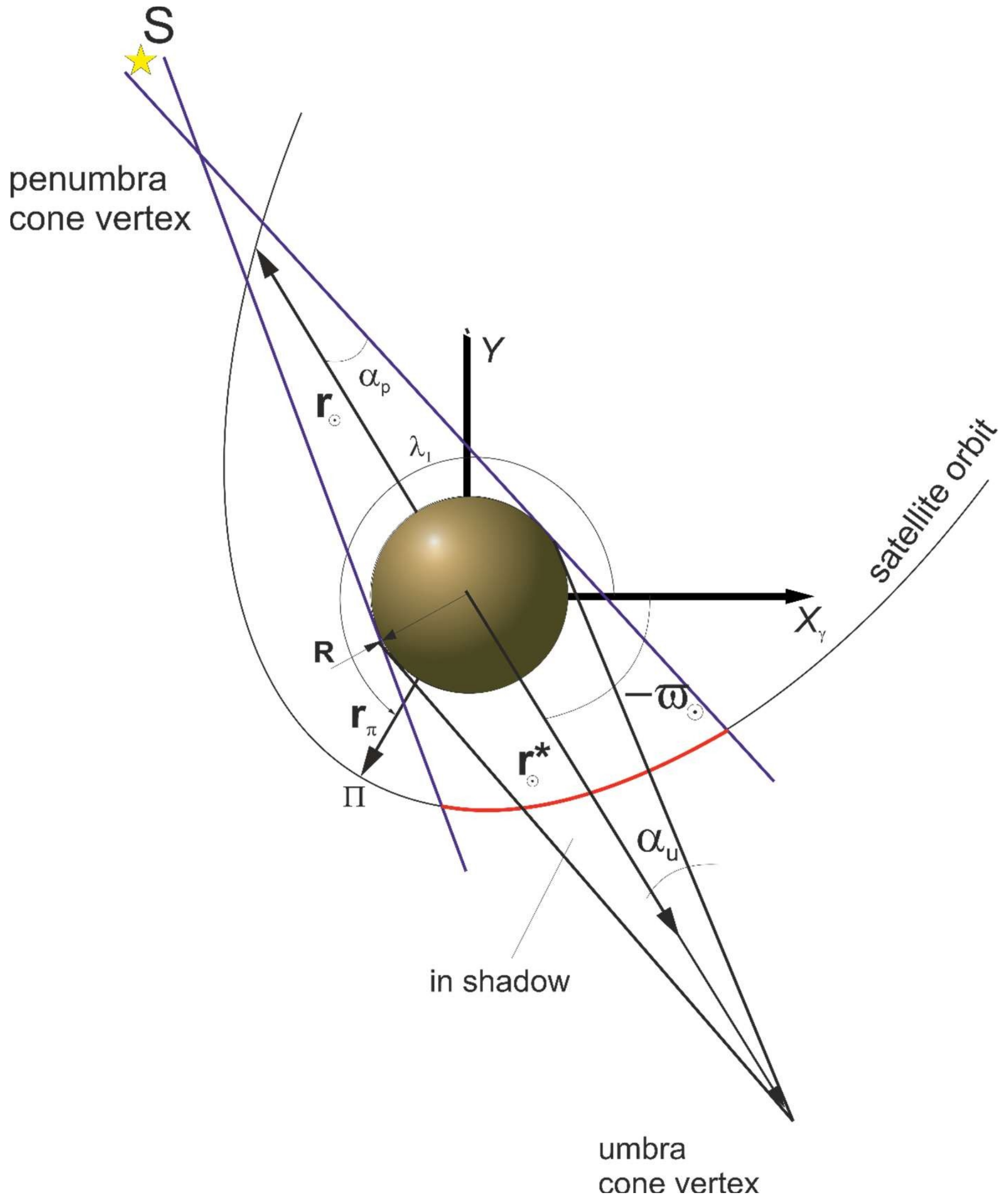

Figure A.1 Calculation scheme for determining the orbital region of the satellite inside the cone of the shadow

Initially, the determination of the satellite's location subsequent to its penumbra exit from the planet is considered. By defining the auxiliary parameter $\gamma$

, which is the angle between the spacecraft planetocentric position and the vector, emanating from the planet's centre of mass and extending towards the direction opposite to the Sun (i.e., $-\mathbf{r}_\odot / \| \mathbf{r}_\odot \|$), the problem can be evaluated. To facilitate problem solving, it is necessary to construct an ancillary triangle (as illustrated in Fig A.2) whose hypotenuse is $r_1 = \| \mathbf{r}_1 \|$; $l$ is aligned towards the direction opposite to the Sun's, and $d$ is transversal to this direction.

From the geometry shown in Fig. A.2, one sees that when the satellite leaves the shadow of the planet, the following equality is satisfied:

(B.1) $$d = \tan\alpha_p \left( l + \frac{R}{\sin\alpha_p} \right),$$

where $l = r_1 \cos\gamma$; $d = r_1 \sin\gamma$ (see Fig. A.1).

Then, the condition of finding the satellite on the boundary of the shadow of the planet may be expressed by the following equation, assuming that the parameters are taken at the moment of the end of the eclipse:

(B.2) $$r_{1,+} \sin\left({}^p\gamma_+\right) = \tan\alpha_p \left( r_{1,+} \cos\left({}^p\gamma_+\right) + \frac{R}{\sin\alpha_p} \right),$$

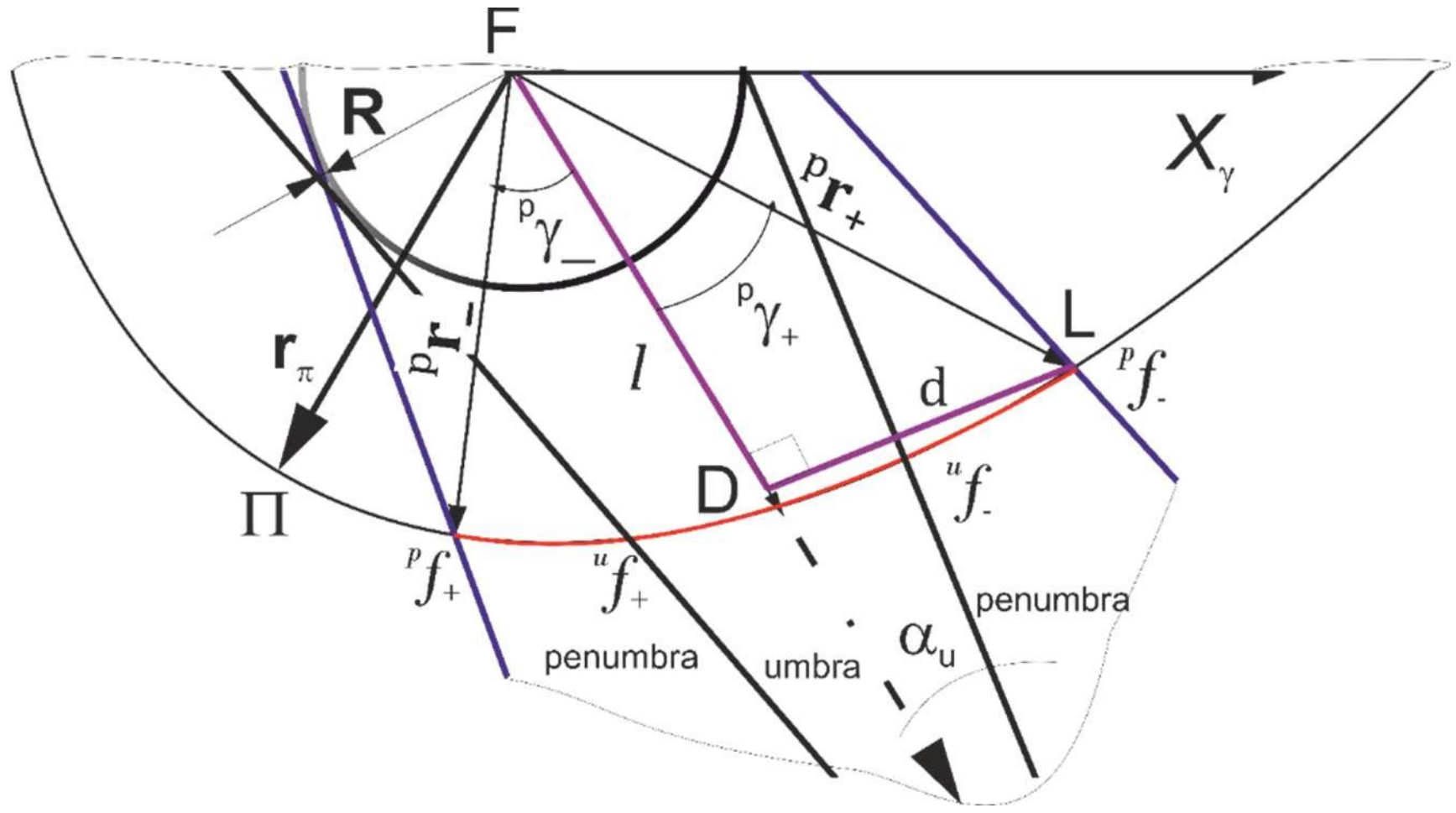

Figure A.2 Calculation scheme of entry/exit positions of satellite to penumbra/umbra cones

Since in equation (B.2) obtained, the parameters ${}^p\beta_+$, $r_{1,+}$ are unknown, to bring it to the equation with only one unknown parameter, one should consider an additional equation:

(B.3) $$r_{1,+} = \frac{p_1}{1 + e_1 \cos(\varpi_\odot + {}^p\gamma_+)}.$$

Then, one can rewrite equation (B.1) as follows:

(B.4) $$\left(p_1 \sin\alpha_p / R\right)(\sin {}^p\gamma_+ / \tan\alpha_p - \cos {}^p\gamma_+) = 1 + e_1(\cos\varpi_\odot \cos {}^p\gamma_+ - \sin\varpi_\odot \sin {}^p\gamma_+).$$

which can then be converted to

(B.5) $$A_1 \sin\left({}^p\gamma_+\right) + B_1 \cos\left({}^p\gamma_+\right) = 1,$$

where $A_1 = \dfrac{p_1 \cos\alpha_p}{R_{pl}} + e_1 \sin\varpi_\odot$, $B_1 = -\left(\dfrac{p_1 \sin\alpha_p}{R_{pl}} + e_1 \cos\varpi_\odot\right)$,

Equation (B.5) is easily solved under the variable $\tilde{v} = \tan\dfrac{{}^p\gamma_+}{2}$:

(B.6) $$\tilde{v}^2 + C_1 \tilde{v} + C_2 = 0,$$

where $C_1 = -\dfrac{2A_1}{B_1+1} = \dfrac{2\left(\mathrm{R}e_1 \sin\varpi_\odot + p_1 \cos\alpha_p\right)}{\mathrm{R}\left(e_1 \cos\varpi_\odot - 1\right) + p_1 \sin\alpha_p}$, $C_2 = -\dfrac{B_1 - 1}{1 + B_1} = -\dfrac{\mathrm{R}\left(e_1 \cos\varpi_\odot + 1\right) + p_1 \sin\alpha_p}{\mathrm{R}\left(e_1 \cos\varpi_\odot - 1\right) + p_1 \sin\alpha_p}$.

By solving equation (B.6) and performing the reverse replacement, one obtains:

(B.7) $${}^p\gamma_{\kappa,+} = 2\arctan\left[\frac{-C_1 \pm_i \sqrt{C_1^2 - 4C_2}}{2}\right], \kappa = 1, 2..$$

Substituting $C_1, C_2$ in the above equation by their values, one obtains the solution for ${}^p\gamma_+$ in the following form considering simplification $\sqrt{1 + \left(-\dfrac{4C_2}{C_1^2}\right)} = \dfrac{1}{2}\left(2 - \dfrac{4C_2}{C_1^2}\right)$. Here and after, because of the linear part of the Taylor series expansion in which the above root was decomposed, the results presented are eligible for eccentricities that do not exceed one unit, i.e., for $e_1 < 1$ orbits.

(B.8) $${}^p\gamma_{1,+} = 2\arctan\left[\frac{1}{2}\left(\frac{1 + e_1 \cos\varpi_S + (p_1/R)\sin\alpha_p}{e_1 \sin\varpi_S + (p_1/R)\cos\alpha_p}\right)\right],$$

(B.9) $${}^p\gamma_{2,+} = 2\arctan\left\{\frac{1}{2}\left[\frac{\begin{array}{l} e_1^2\left[5\cos 2\varpi_\odot - 3\right] - 2 \\ -2e_1\left(p_1/R\right)\left[5\sin(\varpi_\odot - \alpha_p) - 3\sin(\varpi_\odot + \alpha_p)\right] \\ -\left(p_1/R\right)^2\left[5\cos 2\alpha_p + 3\right] \end{array}}{\begin{array}{l} e_1^2 \sin 2\varpi_\odot - 2e_1 \sin 2\varpi_\odot \\ +2e_1\left(p_1/R\right)\cos(\varpi_\odot - \alpha_p) \\ -2\left(p_1/R\right)\cos\alpha_p + \left(p_1/R\right)^2 \sin 2\alpha_p \end{array}}\right]\right\}.$$

It is obvious that ${}^p\gamma_{2,+} > {}^p\gamma_{1,+}$, $\left|{}^p\gamma_{1,+}\right| < 1$; then, the final solution for eclipse end is ${}^p\gamma_{1,+}$, and considering that this algorithm can be applied to find the eclipse onset, the general solution for elliptical orbits may be found as:

(B.10) $${}^p\gamma_{+/-} = 2\arctan\left[\frac{1}{2}\left(\frac{1 + e_1 \cos\varpi_\odot + (p_1/R)\sin\alpha_p}{e_1 \sin\varpi_\odot \pm_\kappa (p_1/R)\cos\alpha_p}\right)\right], \qquad \kappa = +/-.$$

The true anomaly of the satellite at the beginning and end of the eclipse is determined from the following relation:

(B.11) $${}^pf_{+/-} = \varpi_\odot + {}^p\gamma_{+/-}.$$

Additionally, it can be seen that the above equation can be simplified because $\left| tg \frac{^{p}\gamma_{+}}{2} \right| < 1$ can be replaced by series:

(B.12) $$f_{+/-} = \varpi_S + 2\sum_{k=0}^{\infty} \frac{(-1)^k}{(2)^{2k+1}(2k+1)} \left( \frac{1 + e_1 \cos\varpi_{\odot} + (p_1 / R)\sin\alpha_p}{e_1 \sin\varpi_{\odot} \pm_{\kappa} (p_1 / R)\cos\alpha_p} \right)^{2k+1}, \quad \kappa = +/-, \quad e_1 < 1$$

The time of the satellite's eclipses can be calculated using eq. (16).

Note that since the problem has symmetry relative to the direction perpendicular to the sun-planet line, the points of entry/exit to/from the umbra can be found by replacing $\alpha_p$ with $\alpha_u$ in equation (B.1). In this case, one will have a reverse picture when the equation for $d$ can be written as:

(B.14) $$d = \tan\alpha_u \left( -l + \frac{R}{\sin\alpha_u} \right)$$

where the change in sign before $l$ occurs since the umbra cone vertex is located in the direction that is opposite to the Sun.

Repeating the steps for obtaining $^{u}\gamma_{+}$ for the umbra case, a similar equation to the previous case is obtained:

(B.15) $$A_1 \sin\left(^{u}\gamma_{+}\right) + B_1 \cos\left(^{u}\gamma_{+}\right) = 1,$$

where $A_1 = \frac{p_1 \cos\alpha_p}{R} + e_1 \sin\varpi_{\odot}$, $B_1 = \left( \frac{p_1 \sin\alpha_p}{R_{pl}} - e_1 \cos\varpi_S \right)$,

Then, the quadric equation can be received under the variable $\tilde{v} = \tan\frac{^{u}\gamma_{+}}{2}$:

(B.16) $$\tilde{v}^2 + C_1 \tilde{v} + C_2 = 0$$

where $C_1 = -\frac{2A_1}{B_1 + 1} = -\frac{2\left(\mathrm{Re}_1 \sin\varpi_{\odot} + p_1 \cos\alpha_p\right)}{\mathrm{R}\left(e_1 \cos\varpi_{\odot} - 1\right) + p_1 \sin\alpha_p}$, $C_2 = -\frac{B_1 - 1}{1 + B_1} = \frac{\mathrm{R}\left(e_1 \cos\varpi_{\odot} + 1\right) - p_1 \sin\alpha_p}{\mathrm{R}\left(e_1 \cos\varpi_{\odot} - 1\right) - p_1 \sin\alpha_p}$.

By solving equation (B.6) and performing the reverse replacement, the following equation can be obtained:

(B.17) $$^{u}\beta_{+/-} = 2\arctan\left[ \frac{1}{2} \left( \frac{(p_1 / R)\sin\alpha_p - 1 - e_1 \cos\varpi_{\odot}}{e_1 \sin\varpi_{\odot} \pm_{\kappa} (p_1 / R)\cos\alpha_p} \right) \right], \quad \kappa = +/-, e_1 < 1,$$

Note that the above solution is written for the exit/entry point, which is differentiated by the sign "$\pm$" in the denominator. The other root should be neglected since it leads to the wrong one solution, i.e., condition for $^{u}\gamma_{1,+} < \pi / 2$ is violated.